\documentclass{article}              % DO NOT DELETE THIS LINE

\usepackage{biblatex}
\usepackage{xr} % Allows references to labels from another file (e.g. supplementary information)
\usepackage{graphicx}
\usepackage{multirow}
\usepackage{amsmath}
\usepackage{physics}
\usepackage{bm} % Need to use bm package to bold q in the eval/evec subscripts, otherwise it bolds nu, kappa and alpha aswell, no idea why
\usepackage[version=4]{mhchem}  % Adds \ce for chemical formula
\usepackage{url}
\usepackage[capitalise]{cleveref}  % Includes "Fig.", "Eq." etc with \cref. Allows the typesetting of those names to be specified centrally.
\usepackage[export]{adjustbox} % allows \includegraphics[left]
\usepackage[utf8]{inputenc}
\usepackage{fullpage}
\usepackage{listings} % allows use of codeblocks e.g. \begin{lstlisting}[language=Python]
\usepackage{siunitx} % Use \SI command
\usepackage{authblk} % for authors

\DeclareUnicodeCharacter{03BC}{\(\mu\)}
\DeclareUnicodeCharacter{03B1}{\(\alpha\)}

\newcommand{\csubref}[2]{\cref{#1}\textit{(#2)}}
\newcommand{\abinitio}{\emph{ab initio}}
\newcommand{\Abinitio}{\emph{Ab initio}}
\newcommand{\lzo}{\ce{La2Zr2O7}}
\newcommand{\codename}[1]{{\textit{#1}}} % For names of external codes
\newcommand{\code}[1]{{\tt #1}} %For names of classes, functions command-line tools etc.
\newcommand{\filename}[1]{{\tt #1}}
\newcommand{\boldR}{\ensuremath{\mathbf{R}}}
\newcommand{\boldq}{\ensuremath{\mathbf{q}}}
\newcommand{\boldQ}{\ensuremath{\mathbf{Q}}}
\newcommand{\modQ}{\ensuremath{|\boldQ{}|}}
\newcommand{\boldzero}{\ensuremath{\mathbf{0}}}
\newcommand{\boldtau}{\ensuremath{\boldsymbol{\tau}}}
\newcommand{\FCM}{\ensuremath{\Phi(0,l)}} % FCM when referenced in text
\newcommand{\FCMeq}{\ensuremath{\Phi_{\alpha {\alpha}'}^{\kappa {\kappa}'}}} % FCM with super/subscripts for use in equations
\newcommand{\dmateq}{\ensuremath{D_{\alpha {\alpha}'}^{\kappa {\kappa}'}}} % Dynamical mat with super/subscripts for equations
\newcommand{\evec}{\ensuremath{{\mathbf{e}}_{\mathbf{q}\nu\kappa}}}
\newcommand{\eveccmp}{\ensuremath{e_{\mathbf{q}\nu\kappa\alpha}}}
\newcommand{\eveccmpkap}{\ensuremath{e_{\mathbf{q}\nu\kappa^{\prime}\alpha^{\prime}}}}
\newcommand{\eveccmpqvp}{\ensuremath{e_{\mathbf{q^{\prime}}\nu^{\prime}\kappa\alpha}}}

\newcommand{\evecstarcmpqvpbeta}{\ensuremath{e^{*}_{\mathbf{q^{\prime}}\nu^{\prime}\kappa\beta}}}
\newcommand{\freq}{\ensuremath{{\omega}_{\mathbf{q}\nu}}}
\newcommand{\freqprime}{\ensuremath{{\omega}_{\mathbf{q^{\prime}}\nu^{\prime}}}}
\newcommand{\lowerqpt}{\boldq{}-point}
\newcommand{\lowerqpts}{\lowerqpt{}s}
\newcommand{\qpt}{\boldQ{}-point}
\newcommand{\qpts}{\qpt{}s}

\newcommand{\smodqw}{\ensuremath{S(\modQ,\omega)}}
\newcommand{\sqw}{\ensuremath{S(\boldQ, \omega)}}
\newcommand{\qw}{\boldQ{}-\ensuremath{\omega}}
\newcommand{\sfmt}[1]{\textit{(#1)}} % For subfigure caption formatting e.g. \smfmt{a} Si \sfmt{b} Al etc.
\newcommand\D{\partial}  % For partial differential equations
\newcommand{\supp}{supplementary material}

\begin{document}

     %-------------------------------------------------------------------------
     % The introductory (header) part of the paper
     %-------------------------------------------------------------------------

     % The title of the paper. Use \shorttitle to indicate an abbreviated title
     % for use in running heads (you will need to uncomment it).

\title{\codename{Euphonic}: inelastic neutron scattering simulations from force constants and visualisation tools for phonon properties}
\author[a]{Rebecca Fair}
\author[b]{Adam Jackson}
\author[a,c]{David Voneshen}
\author[b]{Dominik Jochym}
\author[a]{Duc Le}
\author[a]{Keith Refson}
\author[a]{Toby Perring}

\affil[a]{ISIS Neutron and Muon Source, STFC Rutherford Appleton Laboratory, Harwell Campus, Didcot, Oxfordshire OX11 0QX, UK}
\affil[b]{Scientific Computing Department, STFC Rutherford Appleton Laboratory, Harwell Campus, Didcot, Oxfordshire OX11 0QX, UK}
\affil[c]{Department of Physics, Royal Holloway University of London, Egham, TW20 0EX, UK}
\date{}
\maketitle
\onecolumn

     % Use \shortauthor to indicate an abbreviated author list for use in
     % running heads (you will need to uncomment it).

%\shortauthor{Soape, Author and Doe}

     % Use \vita if required to give biographical details (for authors of
     % invited review papers only). Uncomment it.

%\vita{Author's biography}

     % Keywords (required for Journal of Synchrotron Radiation only)
     % Use the \keyword macro for each word or phrase, e.g. 
     % \keyword{X-ray diffraction}\keyword{muscle}

\maketitle                        % DO NOT DELETE THIS LINE

\begin{abstract}

Interpretation of vibrational inelastic neutron scattering spectra of complex systems is frequently reliant on accompanying simulations from theoretical models. \Abinitio{} codes can routinely generate force constants, but additional steps are required for direct comparison to experimental spectra. On modern spectrometers this is a computationally expensive task due to the large data volumes collected. In addition, workflows are frequently cumbersome as the simulation software and experimental data analysis software often do not easily interface to each other. Here a new package, \codename{Euphonic}, is presented. \codename{Euphonic} is a robust, easy to use and computationally efficient tool designed to be integrated into experimental software and able to interface directly with the force constant matrix output of \abinitio{} codes.
\end{abstract}

     %-------------------------------------------------------------------------
     % The main body of the paper
     %-------------------------------------------------------------------------
     % Now enter the text of the document in multiple \section's, \subsection's
     % and \subsubsection's as required.

%============================================================================================
%============================================================================================
%           Section 1. Introduction
%============================================================================================
%============================================================================================

\section{Introduction}

The study of atomic vibrations is of both fundamental and applied interest as they have a significant role in many macroscopic material properties. They can drive phase transitions~\cite{budai_metallization_2014}, transport heat~\cite{zheng_advances_2021}, govern elastic properties~\cite{boer_elasticity_2018} and even limit the coherence of qubits~\cite{garlatti_unveiling_2020}. For this reason significant effort has been devoted to understanding atomic vibrations both theoretically and experimentally. Inelastic neutron scattering (INS) is a particularly powerful experimental technique as the momentum and frequency dependence of the atomic vibrations can be straightforwardly and quantitatively related to the experimental intensities, providing a stringent test of theoretical models.

The earliest experiments were performed using a triple-axis spectrometer (TAS) \cite{brockhouse_energy_1955} and this remains a widely applied method. In its conventional form this instrument defines the incident (final) neutron energy with a crystal monochromator (analyser) and a single detector measuring at a single point in \qw{} (momentum-energy) space. Dispersion curves are mapped out by making a series of measurements as a function of energy transfer at fixed momentum, or as a function of momentum at fixed energy transfer, which collectively raster over a \qw{} region of interest. To improve efficiency, additional analyser-detector groups after the sample can be used \cite{kempa_flatcone_2006}, increasing data rates but also increasing the volume and complexity of the data to be modelled.

Alternatively, a time-structured beam may be used in conjunction with beam line components that define the incident neutron energy or final neutron energies, so that each detector in an array can map out many different energy transfers \(\omega\). These time-of-flight (TOF) instruments have become increasingly popular especially with the advent of high power, pulsed spallation sources. To maximise their efficiency these instruments have detector arrays which cover large solid angles broken up into pixels. For example, LET at ISIS \cite{bewley_let_2011} has a three steradian detector array that is split into 98304 pixels, in each of which the energy transfer range is resolved into \(\sim\)300~bins. For single crystal experiments the sample must be rotated, commonly in 0.5\(^{\circ}\)--1\(^{\circ}\) steps over a 180\(^{\circ}\) range, leading to the generation of a multidimensional dataset which consists of \(10^9\)--\(10^{10}\) individual \qw{} points. Data analysis frameworks to handle and interact with these objects exist and are widely used~\cite{ewings_horace_2016,reznik_automating_2020,arnold_mantiddata_2014}.

The calculation of vibrational properties from first-principles or parameterised atomistic lattice dynamics is well established within the computational chemistry and physics communities. As the equations needed are straightforward, many codes have been written to compute INS spectra for comparison between theory and experiment. 
These include \codename{SimPhonies}~\cite{bao_complex_2016}, \codename{OpenPhonon}~\cite{mirone_openphonon_2006}, \codename{Scatter}~\cite{roach_scatter_2007} for \codename{GULP}~\cite{gale_gulp_2005}, \codename{Ab2tds}~\cite{mirone_ab2tds_2013}, the \codename{CLIMAX} series~\cite{kearley_climax_1990,Ramirez-Cuesta2004,cheng_simulation_2019} and \codename{PHONON}~\cite{parlinski_phonon_2014} in addition to private and unreleased codes. Recent versions of \codename{Phonopy} \cite{togo_first_2015} also include functions for calculation of the dynamical structure factor. However no available software meets all of the requirements of a tool suitable for experimentalists at time-of-flight INS facilities, lacking one or more of some functionality such as flexible and customisable modelling of instrumental resolution, the scaling of computational performance to very large, high-resolution datasets, ease of use, and availability of maintained source code. 

The force-constants approach of the above-mentioned codes is restricted to systems and excitations well described by the harmonic approximation of lattice dynamics. Anharmonic phenomena including frequency shifts of asymmetric bond vibrations and overtones, broadening, dynamically stabilized and  high-temperature structural phases and nuclear quantum dynamics are poorly described or not included at all. In some cases it is feasible to still compute stable harmonic phonons using volume or temperature dependent effective potentials \cite{hellman_temperature_2013}. Computational approaches to strong anharmonicity are not as well established or widespread, but one which merits mention in the context of INS spectra is based on analysis of dynamical correlation functions computed using molecular dynamics simulations. The software packages of \codename{nMoldyn} \cite{rog_nmoldyn_2003}, \codename{MDANSE} \cite{goret_mdanse_2017}, \codename{Dynasor} \cite{fransson_span_2021} and \codename{LiquidLib} \cite{walter_liquidlib_2018} all perform the calculation of dynamical structure factor from molecular dynamics trajectory data.

Here we present \codename{Euphonic}~\cite{fair_euphonic_2022}, a software package to calculate inelastic neutron scattering intensities in the harmonic approximation using force constants obtained from \abinitio{} calculations. It is written in Python, for ease of integration with other software and experimental workflows. This allows, for example, it to be used to simulate any part of a time-of-flight \qw{} dataset with instrumental resolution convolution via the data analysis package \codename{Horace} \cite{ewings_horace_2016}. Given the potential size of the \qw{} time-of-flight datasets described above, \codename{Euphonic} has a focus on computational performance and uses an extension written in C and OpenMP to handle the most demanding steps. \codename{Euphonic} is not strongly coupled to any particular atomistic code;
currently, force constants can be read from \codename{CASTEP}~\cite{clark_first_2005}, which internally implements several schemes for phonon calculations, and \codename{Phonopy}~\cite{togo_first_2015}, which manages finite-displacement calculations with any of eleven force calculators.
In addition to simulating large \qw{} time-of-flight datasets, it is intended as a general-purpose tool for analysis of phonon simulations. Command-line programs are included for quick calculation and plotting of phonon band structure, density of states (DOS) and the neutron dynamical structure factor, while the Python API allows \codename{Euphonic} to be used as a more flexible library for optimised phonon frequency and eigenvector calculations and related quantities.

In this paper we will describe: the basic theory behind \codename{Euphonic}; the software structure, its main features, and how it interacts with other software; show how its performance and results compare to similar existing software tools and, finally, compare \codename{Euphonic} to experimental datasets for both single crystals and powders.

%============================================================================================

\subsection{Availability}

\codename{Euphonic} has been developed following software development best practice: continuous integration processes test the functionality and validate numerical results as features are added, and the Python API has been explicitly designed to make it easy to use \codename{Euphonic} with other projects. \codename{Euphonic} is open-source under the GNU General Public License v3 – the source code is is freely available on Github, and packages are distributed by PyPI and Conda-forge for the \codename{pip} and \codename{conda} package managers. Links to the source code and online documentation, including installation instructions and how to use \codename{Euphonic}, are available from the DOI landing page \url{https://doi.org/10.5286/SOFTWARE/EUPHONIC}~\cite{fair_euphonic_2022}.

%============================================================================================

\subsection{Validation datasets}

Throughout this paper a variety of simulated and experimental datasets will be used to compare \codename{Euphonic} outputs to both other software tools and experimental data. The chosen materials are Nb, Al, Si, quartz and \lzo{}, and will be described and used in each section as appropriate. These datasets have been chosen to give a variety of unit cell sizes and crystal symmetries, and include both polar (quartz) and non-polar materials. Simulated force constants have been produced with either \codename{CASTEP} or \codename{VASP}~\cite{kresse_efficient_1996} plus \codename{Phonopy} in order to thoroughly test the features of \codename{Euphonic}. Details of the force constant calculations for each material and where to obtain the results of the simulations and the neutron experimental data can be found in the \supp.

%============================================================================================
%============================================================================================
%           Section 2. Theory
%============================================================================================
%============================================================================================

\section{Theory}

%============================================================================================

\subsection{Harmonic lattice dynamics}

\label{section:harmonic_lattice_dynamics}

 Vibrational excitations --- phonons --- in crystalline materials are described within the theory of lattice dynamics \cite{dove_introduction_1993}.  Displacement of a single atom in an infinite periodic crystal gives rise to forces on every other atom in the crystal, which in the harmonic approximation are linearly related to the displacement by the force constant matrix
\begin{equation}
    \FCMeq(0,l) =
    \frac{\D^{2}E}{{\D}u_{\kappa\alpha0}{\D}u_{{\kappa}'{\alpha}'l}}  
\end{equation}
where \(\Phi\) is the force constant matrix, the unit cell containing the displaced atom is labelled \(0\), \(l\) runs over unit cells in the crystal, \(\kappa\) runs over atoms in the unit cell, \(\alpha\) runs over the Cartesian directions, \(u_{\kappa{\alpha}l}\) is the displacement of atom \(\kappa\) in cell \(l\) in direction \(\alpha\) from its equilibrium position, and \(E\) is the total lattice energy.  These force constants decay rapidly with distance according to a power-law \(|R|^{-n}\) where \(n=5-7\) in non-polar crystals.
This allows for a truncation of \FCM\ at some radius \(R_c\) beyond which the residual force constants may be neglected.  \(R_c\approx{8-10}\)~\AA~ is sufficient for almost all practical cases.

Substituting a plane-wave solution into the equation of motion yields an eigenvalue equation
\begin{equation}
\label{dmat_diag}
    \sum_{\kappa^\prime\alpha^\prime} \dmateq(\boldq)\: \eveccmpkap =
    \freq^{2} \: \eveccmp
\end{equation}
where \dmateq(\boldq) is the
dynamical matrix at wave vector \(\boldq\), the eigenvalues
\(\freq^{2}\) are the square of the phonon frequencies of
mode \(\nu\) at \(\boldq\), and the eigenvectors
\(\evec\) are the phonon polarisation vectors
at \(\boldq\) of mode \(\nu\) for atom \(\kappa\).
The dynamical matrix can be calculated as the mass-weighted Fourier
transform of the force constant matrix
\begin{equation}
\label{dmat_fourier}
    \dmateq(\boldq) =
    \frac{1}{\sqrt{M_\kappa M_{\kappa '}}}
    \sum_{l}\FCMeq{}\:\exp\left(-i\boldq\cdot \boldR_l\right)
\end{equation}
where \(M_\kappa\) is the mass of atom \(\kappa\), \(\boldR_l\) is the vector from the origin to the \(l^\textrm{th}\) unit cell, and \(l\) includes unit cells with \(|\boldR_l| < R_c\).

%============================================================================================

\subsubsection{ Deconvolution of periodic representation of \(\Phi\)}

\label{subsection:periodic_images}
Though the force constant matrix, \FCM, is a finite matrix it
does not map directly onto a modelling framework with supercell
periodic boundary conditions as used in almost all \abinitio{}
lattice dynamics implementations.  Instead these yield a convolution
of \FCM~with the lattice of a suitable supercell. Ideally this would be chosen so the
magnitude of \FCM\ falls to a negligible value in half the
supercell's linear dimension to avoid overlap error. The force constants may either be directly computed
using the supercell as the \abinitio{} simulation cell, (the ``direct
method''), \cite{ackland_practical_1997} or implicitly using Fourier interpolation
either of density-functional perturbation theory \cite{baroni_phonons_2001} or
finite displacements \cite{ackland_practical_1997}. Therefore \codename{Euphonic} must deconvolve a
periodic dataset to recover the aperiodic force constant matrix,
\FCM. For a very large supercell the supercell-periodic images
of  \FCM\ truncated at some radius \(\boldR_c\) do not overlap at all and the extraction
is a straightforward mapping of the data values. However in
practical \abinitio{} calculations the supercell sizes are limited by
computational resources, and there is always a small residual overlap
between periodic images of \FCM.  \codename{Euphonic} adopts the ``Cumulant image''
approach \cite{parlinski_first-principles_1997}, combining the deconvolution with the Fourier
transform to compute the dynamical matrix in a formulation which respects point group symmetry.

%============================================================================================

\subsubsection{Acoustic sum rule}

The invariance of the total energy upon displacement of the entire crystal in space imposes a condition known as the acoustic sum rule which guarantees the presence of the usual three acoustic modes with zero frequency at \(\boldq=\boldzero\) and linear dispersion nearby. The sum rule applies to the force constant matrix or the \(\Gamma\)-point dynamical matrix

\begin{subequations}
 \begin{align}
  \label{eq:asr_realspace}
  \sum_{\kappa l}\FCMeq(0,l) &= 0 \\
  \label{eq:asr_reciprocal}
  \sum_{\kappa} \dmateq(\boldzero) &= 0 \ 
 \end{align}
\end{subequations}

\noindent where \(\kappa\) runs over the \(N\) atoms in the unit cell and \(l\) over the \(N_{\text{cell}}\) unit cells in the supercell. The zeros depend on an exact cancellation of an entire row of the dynamical or force constant matrices, whose elements are large and of opposite sign. In the presence of numerical convergence errors and symmetry breaking by typical \abinitio{} computational grids the sum rule is not exactly satisfied, leading to non-zero acoustic modes at \(\boldq=\boldzero\) and nonlinear dispersion of acoustic modes nearby. 

\codename{Euphonic} optionally applies a correction using projection methods onto the pure translation modes to restore near-exact satisfaction of the sum rule. Two alternative adjustments are implemented; either to the dynamical matrices (the \emph{reciprocal-space} method) or to the periodic representation of the force-constant matrix (the \emph{real-space} method).  If Equation~\ref{eq:asr_reciprocal} is imperfectly satisfied, the dynamical matrix \(D\) (dropping indices for clarity) has three near-zero eigenvalues corresponding to the acoustic modes. A correction is applied
\begin{equation}
 \label{eq:asr_corr_realspace}
  D^{\text{corr}}(\boldq) = D(\boldq) - \Psi \Omega^{\text{acoustic}}\Psi^{-1} 
\end{equation}
where \(\Psi\) is the matrix of eigenvectors of \(D(\boldq=\boldzero)\), \(\Omega^{\text{}acoustic}\) is a matrix whose diagonal entries are the eigenvalues of \(D(\boldq=\boldzero)\) for the acoustic modes and zero otherwise.
As per~\textcite{Gonze1997}, the correction at \(\boldq=\boldzero\) is also applied at non-zero \(\boldq\).
The alternative \emph{real-space} correction is similarly formulated, but applied to the force constant matrix
\begin{equation}
 \Phi^{\text{corr}} =  \Phi - \Pi \Theta^{\text{acoustic}} \Pi^{-1}
\end{equation}
where \(\Phi\), \(\Pi\) and \(\Theta^\text{acoustic}\) are all \(3 N N_{\text{cell}} \times 3 N N_{\text{cell}}\) matrices. \(\Phi\) is the matrix of force constants, \(\Pi\) is the matrix of eigenvectors of \(\Phi\), and \(\Theta^\text{acoustic}\) is a matrix whose diagonal entries are the eigenvalues of \(\Phi\) for the acoustic modes, and zero otherwise. This method more faithfully restores the linearity of the acoustic branches near \(\boldq=\boldzero\) as well as the \(\Gamma\)-point limit.

%============================================================================================

\subsubsection{Polar crystals}

\label{subsection:polar_crystals}

For non-polar crystals, the decay of force constants with distance by a high inverse power law of 5--7 means that a cutoff radius of 8--10~\AA{} is usually sufficient to contain the non-negligible elements of the force constant matrix. The resulting supercell of 16--20~\AA{} in each dimension is well within the typical computational capability of \abinitio{} DFT methods.
(Typically these would be within the local-density approximation or generalised-gradient approximation,
but hybrid functionals have become more accessible even for a system such as \lzo{} \cite{misra_structure_2019}.)
However, in the case of ionic or polar crystals the Coulomb interactions add a tail that decays as \(R^{-3}\), which precludes a converged calculation in a computationally practical supercell. This long-range tail is responsible for the phenomenon of splitting of the longitudinal optical and transverse optical modes (LO/TO splitting) and unwarranted truncation will lead to unphysical behaviour of LO modes at the origin. Fortunately this term may be computed analytically and subtracted from the force constant matrix leaving the remainder term, which can be represented within a feasibly sized supercell~\cite{Gonze1994}. This is the approach adopted by \codename{CASTEP}, \codename{Phonopy} and several other codes, which output only the short-ranged part of the force constant matrix. \codename{Euphonic} computes the dipole-dipole correction term of \cite{Gonze1994} using the Born effective charges and dielectric permittivity to reconstruct the full force constant matrix.

%============================================================================================

\subsection{Coherent inelastic neutron scattering}

The coherent one-phonon scattering structure factors can be calculated at momentum transfer \(\boldQ\) of scattered neutrons \cite{dove_structure_2003,squires_introduction_1996}
\begin{equation}
    \abs{F(\boldQ,\nu)}^2 = \abs{{\sum_{\kappa}{\frac{b_\kappa}{M_{\kappa}^{1/2}\freq^{1/2}}
        (\boldQ\cdot\evec)\:\exp\left(i\boldQ{\cdot}\mathbf{r}_{\kappa}\right)\:\exp\left(-W_{\kappa}\right)}}}^2
    \label{eqn:coh_sfac}
\end{equation}
where \(\boldq\) is the reduced wave vector in the first Brillouin zone (i.e.  \(\boldQ = \boldq + \boldtau\)  where \(\boldtau\) is a reciprocal lattice vector), \(b_\kappa\) is the coherent neutron scattering length of atom \(\kappa\), \(\mathbf{r}_{\kappa}\) is the vector from the origin to atom \(\kappa\) within the unit cell, \(M_\kappa\) is the mass of atom \(\kappa\), \(\evec\) is the phonon polarisation vector at \(\boldq\) of mode \(\nu\) for atom \(\kappa\) and \(\freq\) is the phonon frequency of
mode \(\nu\) at \(\boldq\). The term \(\exp\left(-W_{\kappa}\right)\) is the anisotropic Debye--Waller factor for atom \(\kappa\), where the exponents can be written
\begin{equation}
    \label{eqn:DW_quadratic}
    W_{\kappa} =
        \sum_{\alpha\beta}W_{\kappa}^{\alpha\beta}Q_{\alpha}Q_{\beta}
\end{equation}
\begin{equation}
    \label{eqn:DW_per_atom}
    W_{\kappa}^{\alpha\beta} =
        \frac{\hbar}{4M_{\kappa}N_{q^{\prime}}}
        \sum_{\mathbf{q^{\prime}}\nu^{\prime}\in{BZ}}\frac{\eveccmpqvp\:\evecstarcmpqvpbeta}{\freqprime}
        \mathrm{coth}\left(\frac{\hbar\freqprime}{2k_BT}\right)
\end{equation}
where the sum is over wave vectors and modes in the first Brillouin Zone (BZ), \(N_{q^{\prime}}\) is the number of \lowerqpts{} in the sum, \(T\) is the temperature, \(\alpha\) and \(\beta\) run over the Cartesian directions, \(\hbar\) is the reduced Planck constant and \(k_B\) is the Boltzmann constant. The Debye--Waller factor has been written in this form to make it explicit that the expensive computation of the set of \(3 \times 3 \) matrices \(W_{\kappa}^{\alpha\beta}\) need only be performed once over an appropriately fine grid in the first Brillouin Zone, leaving computation of the Debye--Waller factor for an arbitrary \(\boldQ\) as the fast evaluation of a quadratic form for each atom, \cref{eqn:DW_quadratic}. 

From the one-phonon structure factors the neutron dynamical structure factor \sqw{} can be calculated
\begin{equation}
    \sqw = \frac{1}{2}\sum_{\nu}{\abs{F(\boldQ{},\nu)}^2
        (n_{\mathbf{q}\nu}+\frac{1}{2}\pm\frac{1}{2})
        \delta(\omega\mp\freq)}
    \label{eqn:coh_intensity}
\end{equation}
where the upper and lower signs refer to phonon creation and annihilation respectively and \(n_{\mathbf{q}\nu}\) is the Bose population function
\begin{equation}
     n_{\mathbf{q}\nu} = \frac{1}{\exp\left(\frac{\hbar\freq}{k_{B}T}\right) - 1}
\end{equation}
Finally, the neutron scattering cross-section per unit cell in term of \sqw{} is
\begin{equation}
    \frac{\mathrm{d}^2 \sigma}{\mathrm{d}\Omega\mathrm{d}E_{f}} = \frac{k_{f}}{k_{i}}\sqw{}
\end{equation}
where \(k_{i}\) and \(k_{f}\) are the incident and scattered neutron wave vectors.

%============================================================================================
%============================================================================================
%           Section 3. Implementation
%============================================================================================
%============================================================================================

\section{Implementation}

\label{section:features}
\begin{figure}
\caption{Where \codename{Euphonic} sits in relation to other software, and users of \codename{Euphonic} or those software.}
\scalebox{1.}{\includegraphics{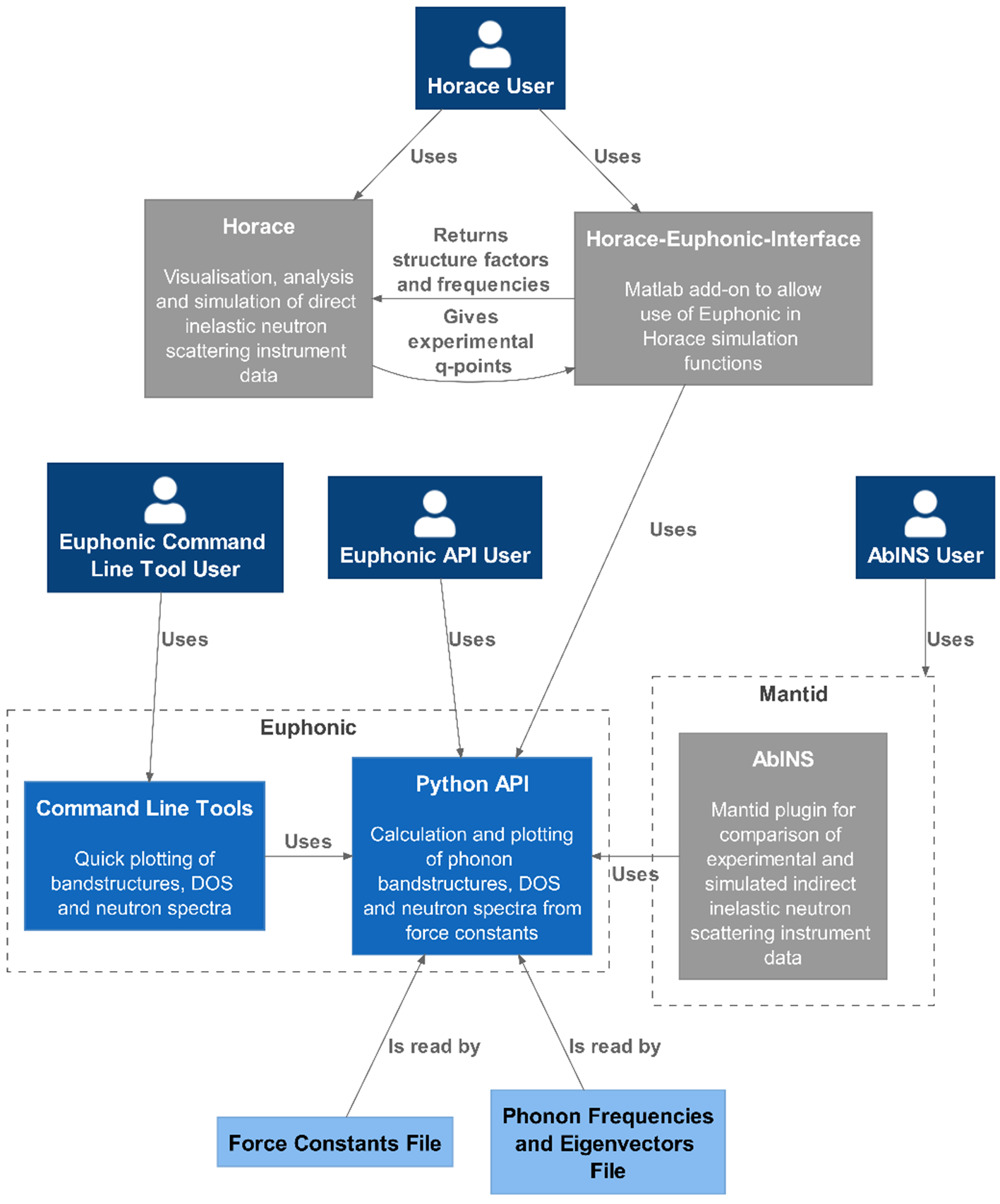}}
\label{fig:euphonic_context}
\end{figure}

\codename{Euphonic} provides an extensive Python API and a number of convenient command line tools. \cref{fig:euphonic_context} shows how \codename{Euphonic} connects with existing packages in the neutron software ecosystem. Established software such as \codename{Horace}~\cite{ewings_horace_2016} and the \codename{Mantid}~\cite{arnold_mantiddata_2014} plug-in \codename{AbINS}~\cite{dymkowski_abins_2018} make direct use of \codename{Euphonic} as a calculator for simulated phonons and scattering intensities. The command line tools provide convenient plotting of phonon band structures, DOS and the neutron dynamical structure factor along specific reciprocal lattice directions (see \cref{section:cl_tools}). \codename{Euphonic} can also be used directly from Python environments for customised plots, workflows or functionality development.

%============================================================================================

\subsection{Context and API}

In a typical \codename{Horace} workflow, the scattering intensities at millions of \qpts{} can be combined to produce multidimensional plots of measured data. Prior to the availability of \codename{Euphonic}, the workflow for generating such plots was limited by the need to read pre-computed phonon frequencies and eigenvectors from files produced by other programs. The bottleneck here becomes the activity of reading and writing to disk. For a system of \(N\) atoms per unit cell, each \qpt{} has a set of eigenvectors whose storage requirements of \(18N^2\) floating-point numbers can become impractically large. For example, a 22-atom unit cell with a modest 25,000 \qpts{} would equate to a 5~GB text file in \codename{CASTEP} \filename{.phonon} format. In a typical cluster environment running the \codename{CASTEP} \code{phonons} tool, more than 85\% of run time is spent writing this file to disk (see \cref{table:castep_phonons_write} for specific \(N^2\) timing examples). Furthermore, on a shared cluster resource this can be exacerbated by local network capacity and the activities of other users.

\begin{figure}
\caption{A summary of the API to \codename{Euphonic}, showing the main classes and methods.}
\scalebox{1.}{\includegraphics{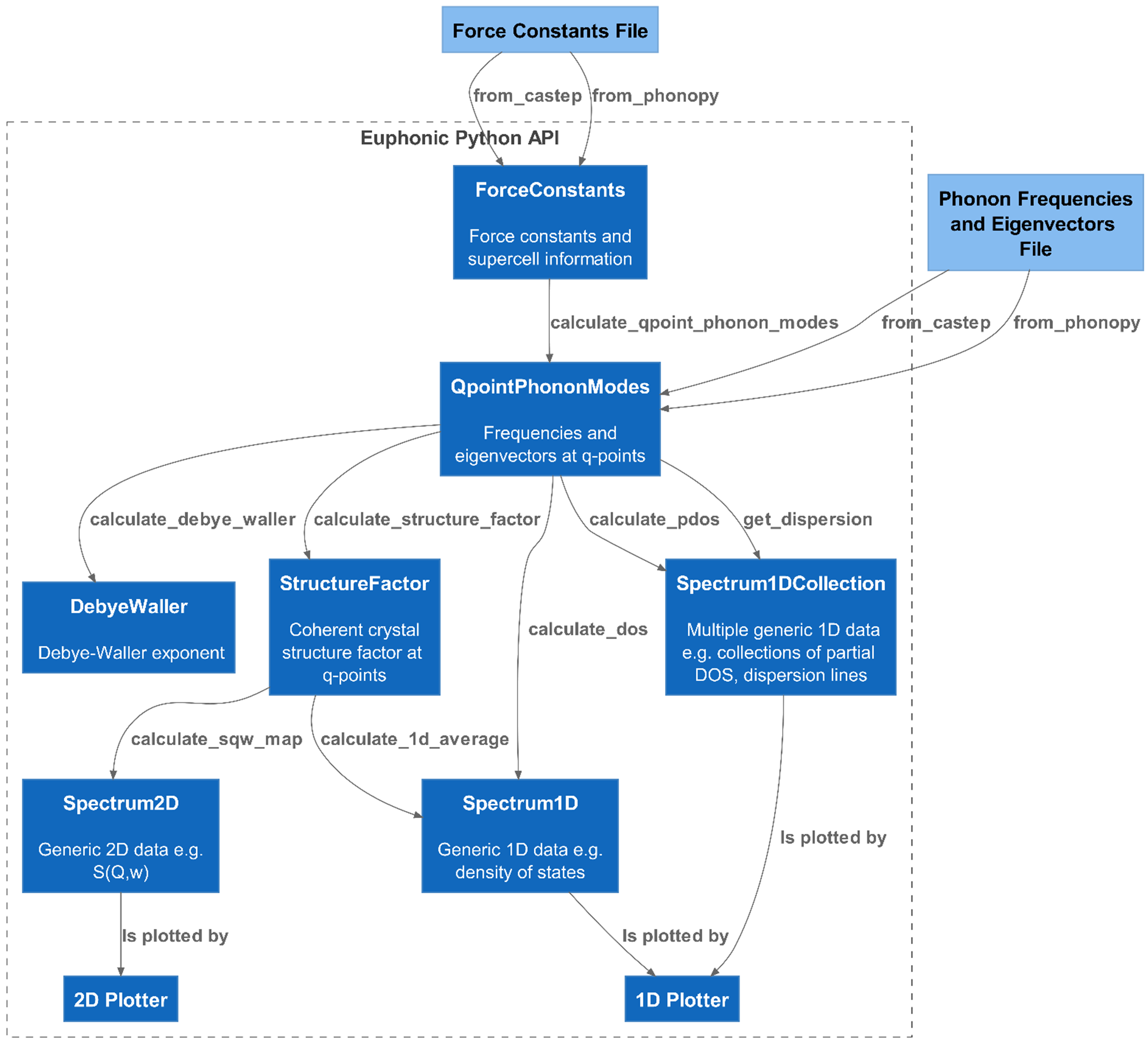}}
\label{fig:euphonic_api}
\end{figure}

The requirement of efficiency for \codename{Euphonic} has driven the decision to implement Fourier interpolation of phonon frequencies and eigenvectors directly from force constants. This allows for calculation of data for each \qpt{} on demand, removing any need for file-based data transfer from other codes. Accordingly, the representation of force constants as a class and associated methods forms the core of the \codename{Euphonic} API, illustrated in \cref{fig:euphonic_api} and described for reference below.

\code{ForceConstants} objects can be instantiated from Python data objects such as \codename{Numpy}~\cite{harris_array_2020} arrays, but would typically
be created from the data outputs of an external modelling code --
currently \codename{CASTEP} \filename{.castep\_bin} output and \codename{Phonopy} \filename{phonopy.yaml}, \filename{FORCE\_CONSTANTS} and \filename{force\_constants.hdf5} output are supported. Through \codename{Phonopy}, \codename{Euphonic} force constants can be obtained using a wide variety of atomistic codes such as \codename{VASP}~\cite{kresse_efficient_1996}, \codename{Abinit}~\cite{gonze_abinit_2020} and \codename{Quantum Espresso}~\cite{giannozzi_quantum_2009}, making \codename{Euphonic} accessible to a large portion of the materials modelling community.
From the force constants, phonon frequencies and eigenvectors may be calculated at arbitrary \qpts{} using the methods described in \cref{section:harmonic_lattice_dynamics}.

\code{QpointPhononModes} represents phonon mode data: the \qpts{}, phonon frequencies and eigenvectors. \code{QpointPhononModes} can also be instantiated from external data files (currently, these are \codename{CASTEP} \filename{.phonon} and \codename{Phonopy} \filename{mesh/band/qpoints} files).
The \codename{Pint} Python library~\cite{grecco_pint_2012} is extensively used in \codename{Euphonic} to represent dimensioned data as a \code{Quantity} with both a magnitude and a unit. This makes the units explicit, and facilitates conversion to the end user's preferred units, usually one of meV, cm\(^{-1}\) or THz, in the case of phonon frequencies.
No particular \qpt{} sampling is enforced; while, for example, a Monkhorst--Pack mesh is recommended for traditional DOS plotting, phonon modes can also be calculated along a high-symmetry path or sampled at random points depending on the use case. 
From the phonon modes, \codename{Euphonic} can compute quantities such as the mode-resolved structure factors, the Debye--Waller exponent and total, partial and neutron-weighted DOS.

\code{StructureFactor} is derived from the phonon modes. This object contains the neutron structure factors resolved by \qpt{} and phonon mode index (\(\nu\)) and also the \qpts{} and frequencies. This can be binned in energy to produce a two-dimensional (2D) \sqw{} plot, or averaged over the contained \qpts{} to produce a one-dimensional (1D) \(S(\omega)\) spectrum.

\code{DebyeWaller} represents the temperature dependent Debye--Waller factor; specifically, it contains the set of $3 \times 3$ matrices, one per atom, \(W_{\kappa}^{\alpha\beta}\) defined in \cref{eqn:DW_per_atom}. A \code{DebyeWaller} object is typically pre-computed over a uniform \qpt{} mesh and then applied during a structure factor calculation to perform the fast calculation of the atomic Debye--Waller exponent \(W_{\kappa}\) via \cref{eqn:DW_quadratic} over whichever (potentially large) set of \qpts{} are appropriate to the task at hand.

\code{Spectrum2D}, \code{Spectrum1D} and \code{Spectrum1DCollection} are classes representing generic spectrum objects that can be used for various purposes, such as representing the DOS with \code{Spectrum1D}, or a \sqw{} or \smodqw{} intensity map with \code{Spectrum2D}. Band structure and partial DOS data are represented with \code{Spectrum1DCollection}, which ensures consistent bins are used and allows individual lines to be tagged with metadata. The plotting tools in turn work with the generic spectrum objects to produce plots of phonon band structure, DOS, and intensity maps.

\code{Crystal} is a simple class containing the crystal structure information: the cell vectors, atom positions, species and masses. The above \code{ForceConstants}, \code{QpointPhononModes}, \code{StructureFactor} and \code{DebyeWaller} classes all contain an instance of this crystal class as an attribute, to ensure the data in each class remains complete and unambiguous.

%============================================================================================

\subsection{Use with \codename{Horace}}

\label{section:use_with_horace}

A widely used software application for analysis and visualisation of multidimensional time-of-flight inelastic neutron scattering data from single crystal experiments is \codename{Horace} \cite{ewings_horace_2016}. In addition to handling and plotting the data, it also allows simulation and fitting of these datasets with user-created models of the scattering function. A Matlab add-on has been developed, \codename{Horace-Euphonic-Interface} \cite{fair_horace-euphonic-interface_2022} which provides interface functions that allow \codename{Euphonic} to be used to simulate datasets directly in \codename{Horace}. The way this works is illustrated in \cref{fig:euphonic_context}. First a user sets up a model with \codename{Horace-Euphonic-Interface}, giving the path to the force constants file or folder, and adding other optional parameters such as the sample temperature or the Debye--Waller grid size. The user then calls a \codename{Horace} simulation function with the dataset to be simulated and the model they have just created. \codename{Horace} will automatically provide the \qpts{} to be simulated to \codename{Horace-Euphonic-Interface}, which then calls \codename{Euphonic} to calculate the structure factors and phonon frequencies at those \qpts{}. \codename{Horace-Euphonic-Interface} then converts the output from \codename{Euphonic} into the required form for \codename{Horace} to create the simulated dataset.

This is a significant improvement over previous workflows to simulate scattering from phonons, which required users to program their own functions to calculate the structure factors from \abinitio{} calculations. These would not usually include Fourier interpolation of the force constants, and hence were restricted to the \qpts{} output by the \abinitio{} calculations. These bespoke scripts were also not typically optimised for fast computation. \codename{Horace-Euphonic-Interface} has allowed users to easily simulate on exactly the same axes and in the same software as the experimental data, making fitting of scaling factors and quantitative comparisons much more convenient. The performance of \codename{Euphonic} has also made application of the Monte Carlo based instrumental resolution convolution method~\cite{perring_high_1991} available in \codename{Horace} tractable with phonons for the first time. Examples of data simulated and fitted with \codename{Euphonic} and \codename{Horace} are shown in \cref{section:experimental_data_crystal}. \codename{Horace-Euphonic-Interface} is open source and distributed as a Matlab Toolbox file on Github and the Matlab File Exchange. Links to the source code and online documentation are available from the DOI landing page \url{https://doi.org/10.5286/SOFTWARE/HORACEEUPHONICINTERFACE}~\cite{fair_horace-euphonic-interface_2022}.

%============================================================================================

\subsection{Use with \codename{AbINS}}

\label{section:use_with_abins}

\codename{AbINS} is a code which simulates powder-averaged INS spectra in an analytic incoherent approximation, and resides in the \codename{Mantid} framework used for experimental data reduction and analysis~\cite{dymkowski_abins_2018,arnold_mantiddata_2014,akeroyd_mantid_2013}.
\codename{AbINS} has recently been updated to make use of \codename{Euphonic}; as of \codename{Mantid} version 6.3,
it is possible to select force constants data in \codename{CASTEP} or  \codename{Phonopy} format as an input file.
End users do not need to know anything about \codename{Euphonic} or change their workflow,
except to ensure that force constants data is present in their \filename{.castep\_bin} or \filename{phonopy.yaml} files.
A single-parameter cutoff distance (as defined by \textcite{moreno_optimal_1992}, and hidden from the user interface) is used
to determine a default Monkhorst--Pack mesh for the given crystal structure;
eigenvalues and eigenvectors are computed using the \codename{Euphonic} Python API and passed on to the usual incoherent inelastic scattering computation \cite{dymkowski_abins_2018}.
\codename{AbINS} and \codename{Horace} have different target users: with \codename{Euphonic} as a common dependency, it becomes possible for these neutron-scattering simulation codes to develop overlapping feature sets while sharing implementation work.

%============================================================================================
%============================================================================================
%           Section 4. Performance profiling and optimisation
%============================================================================================
%============================================================================================

\section{Performance profiling and optimisation}

\label{section:performance}

Given the aim of calculating scattering intensities at millions of \qpts{}, performance optimisation of \codename{Euphonic} has been a priority. \cref{table:castep_phonons_write} illustrates the potential cost of writing large eigenvector arrays to disk, which has been avoided in \codename{Euphonic} by enabling its own interpolation from force constants. There are four main steps in computing the neutron dynamical structure factor from force constants: reading the force constants, computing the phonon frequencies and eigenvectors (interpolation), computing the mode-resolved structure factors and, finally, applying the Bose factor and binning these in energy to obtain \sqw{}. Reading the force constants only has to be done once, and the binning for large datasets is typically done via another program such as \codename{Horace}.

Of the two remaining parts, the calculation of phonon frequencies and eigenvectors is by far the most expensive, illustrated in \cref{table:compare_interpolate_sf}. Even in the case of \lzo{}, which has the most expensive structure factor calculation due to the number of atoms in the unit cell, the interpolation takes approximately 20 times longer. For this reason, this part of the calculation has been the focus of much of the optimisation effort, and performance comparisons have been made with another interpolation tool, the \codename{CASTEP} \code{phonons} tool, rather than software that performs the cheaper structure factor calculations such as \codename{Ab2tds} \cite{mirone_ab2tds_2013} or \codename{OClimax} \cite{cheng_simulation_2019,cheng_calculation_2020}. The \codename{Phonopy} software does perform phonon interpolation, but has not been chosen for performance comparisons as it does not (as of version 2.11.0) parallelise its calculation over \qpts{}.

\begin{figure}
\caption{Comparison of the wall time taken to run the \code{calculate\_qpoint\_phonon\_modes} interpolation function in \codename{Euphonic} against the \code{phonon\_calculate} function in \codename{CASTEP} for 25,000 \qpts{} for different materials and numbers of processors. Scatter points show the wall time taken to run the serial Python version of \code{calculate\_qpoint\_phonon\_modes}. (Details of the hardware are given in \cref{section:hardware_and_software}.)}
\scalebox{1.}{\includegraphics{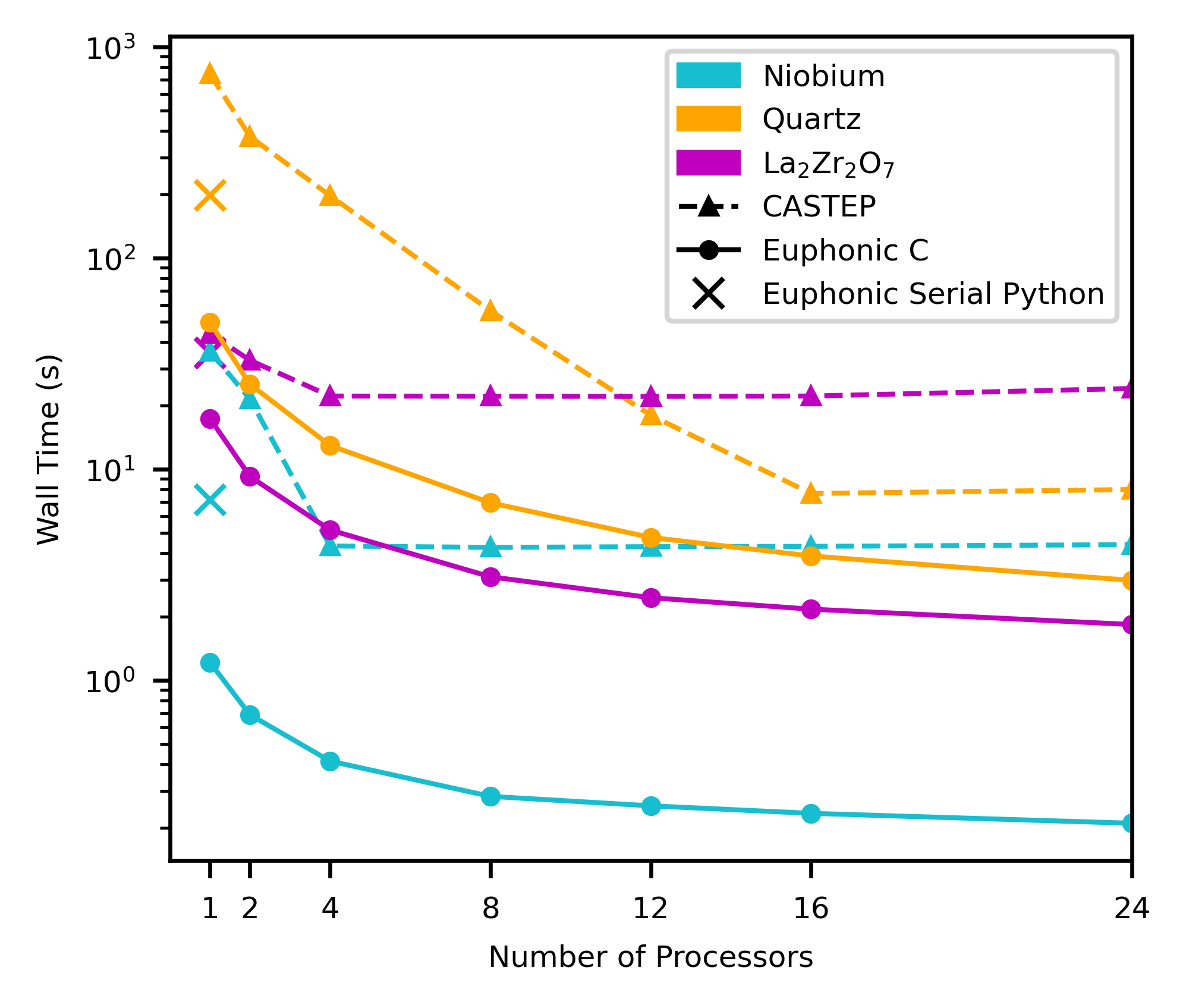}}
\label{fig:walltime_compare}
\end{figure}

\codename{Euphonic} makes extensive use of \codename{Numpy} to improve its performance, but the serial Python performance shown in \cref{table:compare_interpolate_sf} is still not sufficient to simulate the numbers of \qpts{} contained in larger multidimensional \qw{} datasets in a reasonable time. Accordingly, an extension has been written in C and OpenMP to perform the interpolation part of the calculation, which both improves performance significantly and enables calculations to be carried out in parallel.
The performance improvement can be seen in \cref{fig:walltime_compare}, which shows the time taken to run the \code{calculate\_qpoint\_phonon\_modes} interpolation function in \codename{Euphonic} for 25,000 \qpts{} for different materials, for both the serial Python and parallel C implementations. One metric for comparing performance is the speedup calculated as the ratio of the times to perform the same operation in the two implementations:
\begin{equation}
\label{eq:speedup}
    S = \frac{T_1}{T_2}
\end{equation}
For one processor, use of the C extension gives speedups of 2.1, 4.0 and 6.0 over the pure Python implementation for \lzo{}, quartz and Nb respectively.
\cref{fig:walltime_compare} also shows how the wall time changes with increasing numbers of processors and compares it with the time taken to run the \code{phonon\_calculate} function from the \codename{CASTEP} \code{phonons} tool. Care has been taken to get a fair comparison, so \codename{CASTEP} features that are not available in \codename{Euphonic} which would have decreased the performance of \codename{CASTEP} have been switched off (group theory analysis, dynamical matrix symmetrisation) and features that could not be turned off have been profiled and subtracted from the total time (writing the \filename{.phonon} file, constructing the force constant matrix). Even with this taken into account, the performance of \codename{Euphonic} is better than the \codename{CASTEP} \code{phonons} tool by an order of magnitude in some cases, with \codename{Euphonic} giving speedups of 13.1, 2.7 and 20.9 over the \codename{CASTEP} tool for \lzo{}, quartz and Nb respectively for 24 processors.

It can be seen in \cref{fig:walltime_compare} that interpolation for quartz is slower than \lzo{}, despite having fewer atoms per unit cell. This is due to the expensive Ewald sum correction that must be applied to the dynamical matrix for polar materials, described in \cref{section:harmonic_lattice_dynamics}. This calculation has been heavily optimised in \codename{Euphonic}, and explains the performance difference between \codename{Euphonic} and the \codename{CASTEP} \code{phonons} tool for quartz. In the case of Nb and \lzo{}, the performance discrepancy largely comes from the fact that the \codename{CASTEP} \code{phonons} tool uses distributed memory parallelism via MPI, so needs to communicate the phonon frequencies and eigenvectors back to the main process, which is where most time is spent. \codename{Euphonic} does not have this issue as it makes use of shared memory parallelism via OpenMP. This was chosen specifically to satisfy the two main use cases for \codename{Euphonic}: running smaller calculations on a single computer or node; and running larger calculations on a cluster via a data analysis tool such as \codename{Horace}, in which case \codename{Horace} would handle any multi-node parallelism.

\begin{figure}
\caption{Speedup of the \code{calculate\_qpoint\_phonon\_modes} interpolation function in \codename{Euphonic} with C extension compared to serial Python for 250,000 \qpts{} for different materials and numbers of processors. (Details of the hardware are given in \cref{section:hardware_and_software}.)}
\scalebox{1.}{\includegraphics{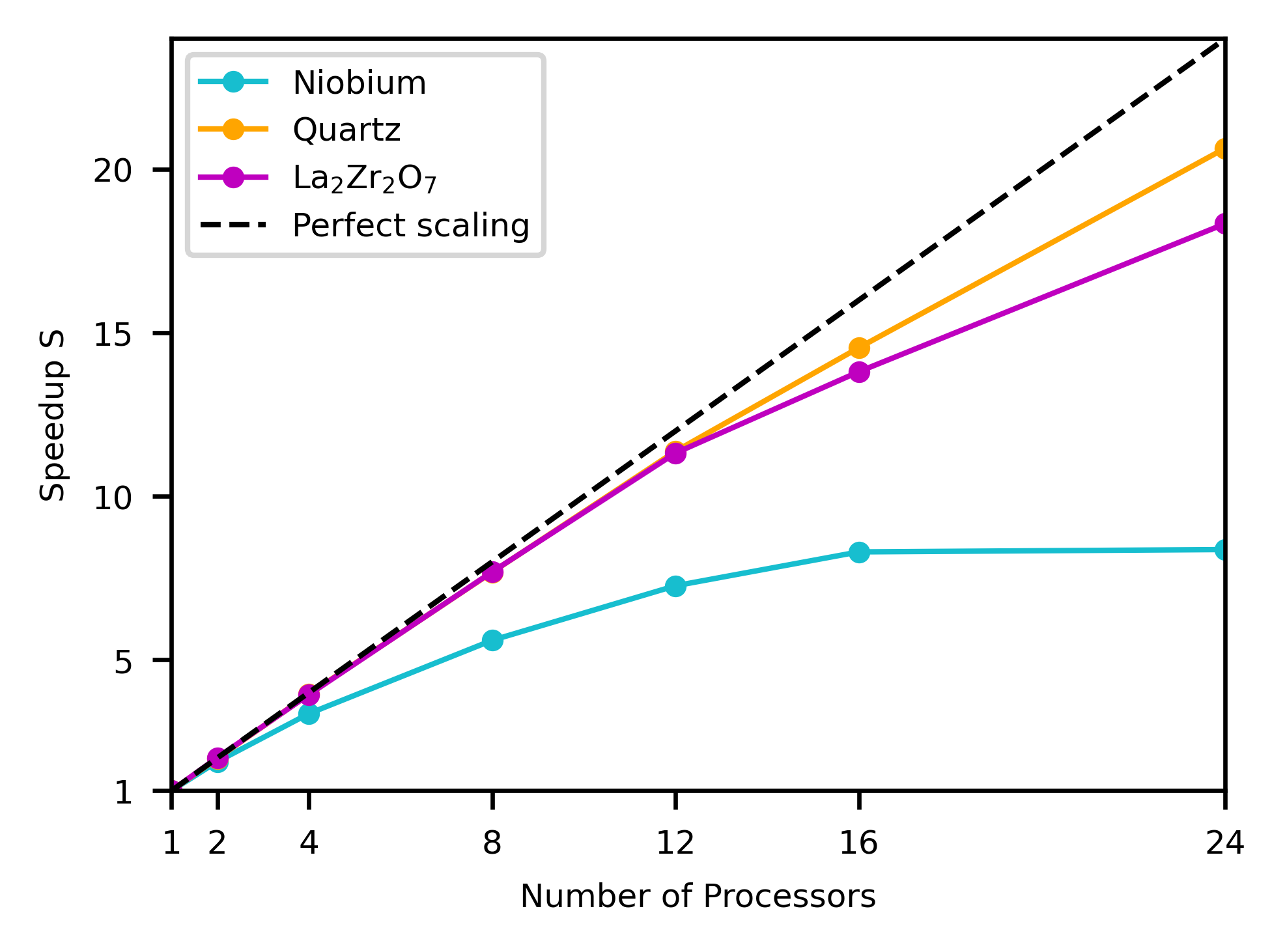}}
\label{fig:euphonic_scaling_250k}
\end{figure}

\begin{figure}
\caption{Where time is spent in the \codename{Euphonic} interpolation function \code{calculate\_qpoint\_phonon\_modes} for 250,000 \qpts{} for different materials and numbers of processors. The white areas indicate time spent in the serial part of the code.}
\scalebox{1.}{\includegraphics{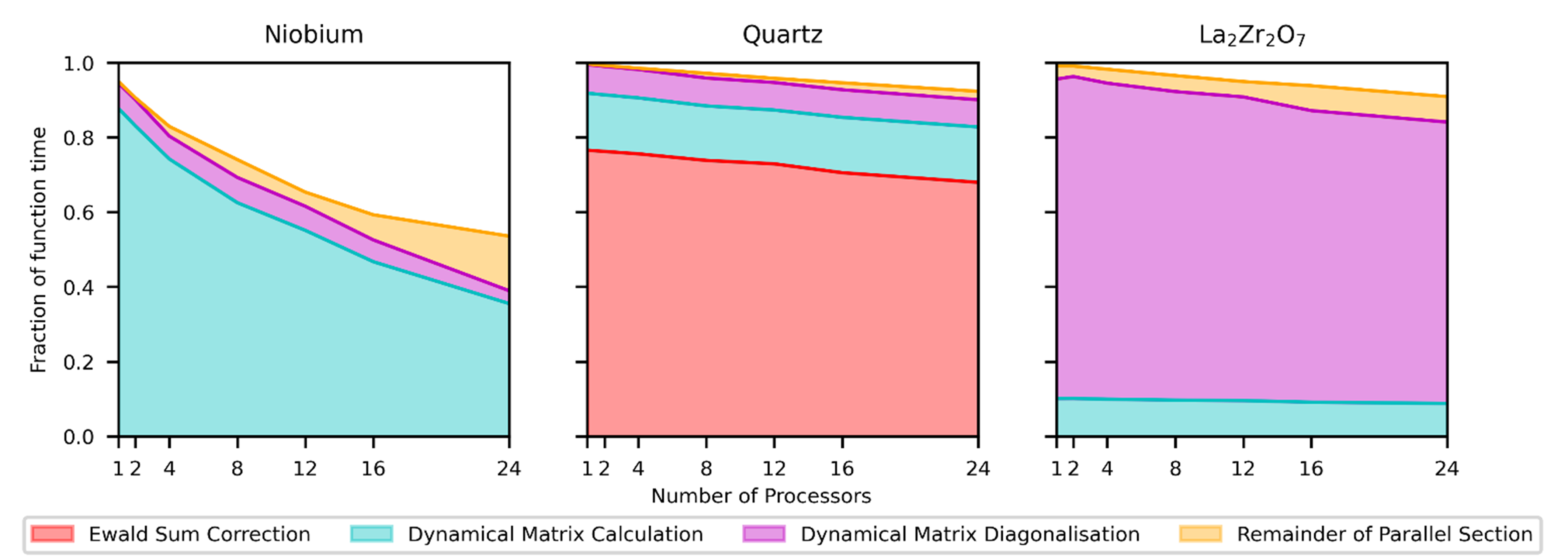}}
\label{fig:cext_prof}
\end{figure}

The number of \qpts{} used for comparison of \codename{Euphonic} to other tools, 25,000 for the benchmarking presented above, was chosen because it pushes the limits of the \codename{CASTEP} tool, and allows results to be obtained in a reasonable amount of time. However, the run times of just a few seconds for \codename{Euphonic} for large numbers of processors are not enough to obtain good performance data, as a non-negligible amount of time will be spent in non-computational parts of the code, for example importing libraries or spawning threads. The interpolation in \codename{Euphonic} has therefore also been profiled for 250,000 \qpts{}, and has been used to demonstrate the performance scaling, this is shown in \cref{fig:euphonic_scaling_250k}. The speedup has been calculated as in \cref{eq:speedup}, where \(T_1\) is the serial interpolation function time, and \(T_2\) is the parallel time. Each interpolated \qpt{} is independent, so the calculation can easily be parallelised over \qpts{} using a parallel \code{for} loop. Despite this independence of \qpts{}, the scaling is imperfect, particularly for Nb. This can be explained by looking at \cref{fig:cext_prof}, which shows the time spent in different parts of the interpolation calculation in C for different numbers of processors and materials. In particular, the serial Python part of the calculation shown in white imposes a \(\sim\)0.1~s overhead, which limits the maximum possible speedup, especially for a small system like Nb where the parallelised part only takes 0.1~s with 24 processors. The serial part includes various set-up tasks, such as calculating the list of periodic supercell images as described in \cref{subsection:periodic_images}.

\cref{fig:cext_prof} also explains the longer run times for quartz seen in \cref{fig:walltime_compare}. Even after being the focus of much optimisation work, the Ewald sum still takes around 70\% of the total interpolation time for quartz. This optimisation has included avoiding redundant calculation of \boldQ-independent values by factorisation, and optimising the balance between the real and reciprocal space sums (\(\Lambda\) in equation 5 in \textcite{Gonze1994}). Changing \(\Lambda\) will not change the result but can drastically improve the performance if chosen correctly. The optimum value depends on the material and is not immediately obvious --- \codename{Euphonic} provides a command-line tool, \code{euphonic-optimise-dipole-parameter}, which profiles the calculation for a few \qpts{} and suggests the optimum value for that system for use in further calculations. Even with these optimisations, there is still a performance penalty for the calculation of phonons for polar materials, suggesting a target for further optimisation work. In non-polar materials, most of the time is spent either calculating or diagonalising the dynamical matrix. This depends strongly on the number of atoms in the unit cell vs. the number of cells in the supercell.
In the case of \lzo{}, which has 22 atoms in the unit cell, over 70\% of the interpolation time is spent diagonalising the dynamical matrices. By contrast Nb has 1 atom per unit cell and (ignoring serial overhead) over 80\% of the time is spent calculating dynamical matrices.

%============================================================================================

\subsection{Hardware and software libraries}

\label{section:hardware_and_software}

All profiling in this section has been completed on the STFC Scientific Computing Department’s SCARF Cluster using the SCARF 18 hardware, which contains 2 Intel Gold 6126 processors per node (24 cores per node). \codename{Euphonic} 0.6.1 and \codename{CASTEP} 19.1 were used. Both the \codename{Euphonic} C extension and the \codename{CASTEP} \code{phonons} tool have been compiled with the Intel 2018.3 compiler, OpenMPI 3.1.1 and linked against Intel MKL 2018.3. The profiling results can be obtained at \url{https://github.com/pace-neutrons/euphonic-performance}.

%============================================================================================
%============================================================================================
%           Section 5. Validation
%============================================================================================
%============================================================================================

\section{Validation}

The results of a comparison of \codename{Euphonic} output to experimental inelastic neutron scattering data will be given in \cref{section:experimental_data_crystal}. In this Section, to test the calculation of the dynamical structure factor stringently, \codename{Euphonic} is validated against two other programs - \codename{Ab2tds} \cite{mirone_ab2tds_2013} and \codename{OClimax} \cite{cheng_simulation_2019,cheng_calculation_2020}. This allows validation of the more subtle parts of the calculation (such as the Debye--Waller factor) without the complication of other scattering mechanisms or broadening due to the instrumental resolution, nor the possibility of the first-principles computation of the force constant matrix failing to fully capture the physics of the lattice dynamics. For the validation comparisons, four materials have been chosen: \lzo{}, quartz, Nb and Al. \lzo{}, quartz and Nb force constants,  frequencies and eigenvectors have been computed using \codename{CASTEP}, and the corresponding  data for Al has been computed using \codename{VASP} and \codename{Phonopy}, to validate both the \codename{CASTEP} and \codename{Phonopy} readers available in \codename{Euphonic}.

\begin{figure}
\caption{The neutron dynamical structure factors for the validation cuts simulated with \codename{Euphonic}. Nb along \sfmt{a} \([h,h,0]\), \sfmt{b} \([2-k,k,0]\). Quartz along \sfmt{c} \([h,-4,0]\), \sfmt{d} \([-3,0,-l]\). \lzo{} along \sfmt{e} \([-5,7,-l]\), \sfmt{f} \([h,-h,2]\). Al along \sfmt{g} \([h,2,2]\), \sfmt{h} \([h,2+\frac{h}{2},2+\frac{h}{2}]\).}
\scalebox{1.}{\includegraphics{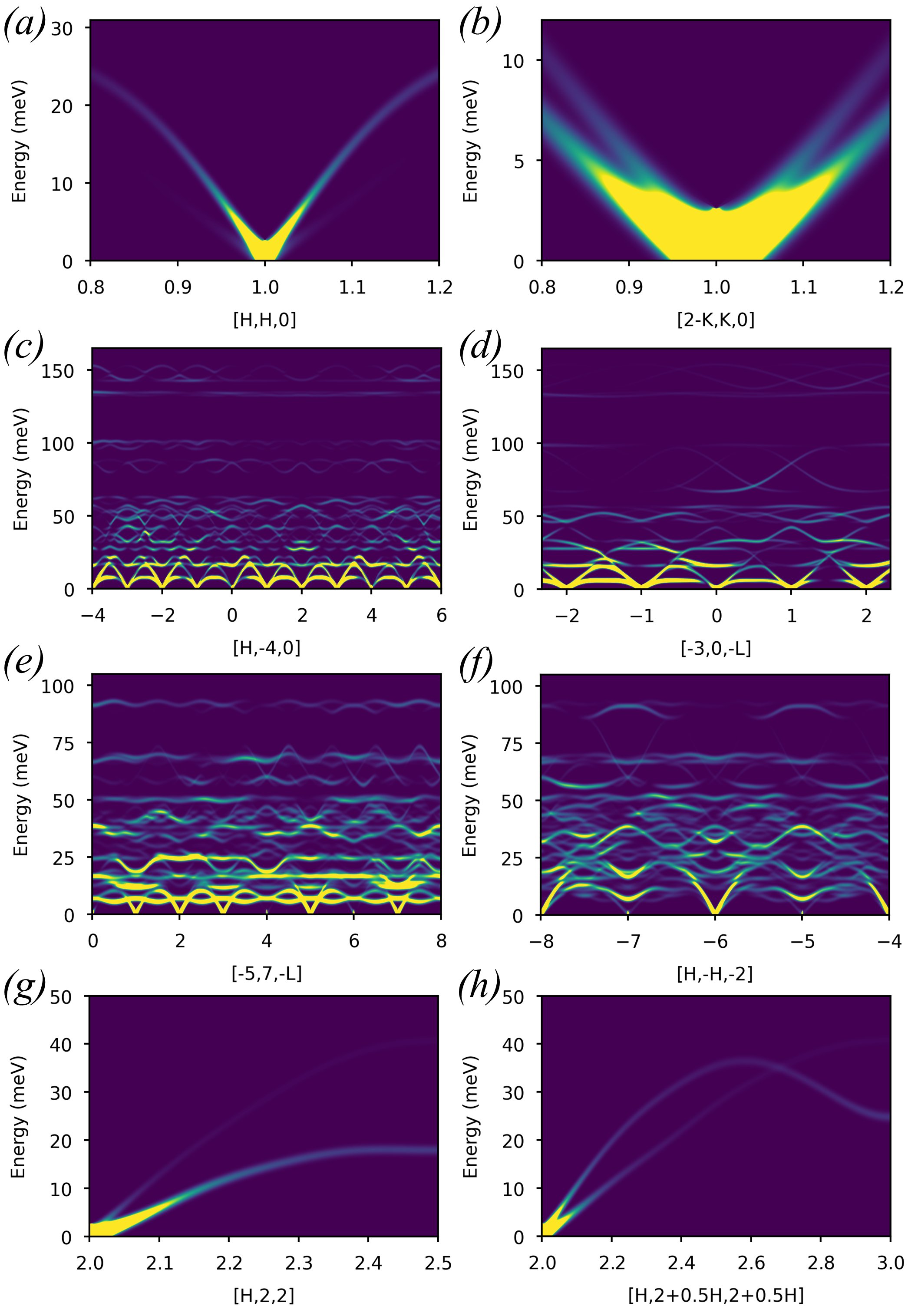}}
\label{fig:validation_cuts}
\end{figure}

The data chosen for validation are all 2D (\boldQ,\(\omega\)) maps as these types of data can be calculated by all three programs. The data maps have a wide variation of magnitudes and directions in \boldQ{} to ensure the variation of quantities such as structure factors and Debye--Waller factors across \boldQ-space are reliably tested. Visualisations of the chosen (\boldQ,\(\omega\)) maps are shown in \cref{fig:validation_cuts}. The metric that has been used for comparison is the mean relative percentage difference (MRPD):
\begin{equation}
\label{eq:mrpd}
    \frac{1}{n}\sum_{i=1}^{n}\frac{|y_i-x_i|}{x_i}\times100
\end{equation}
\noindent where \(x_i\) is the neutron dynamical structure factor (\sqw{} as in \cref{eqn:coh_intensity}) calculated with \codename{Euphonic}, and \(y_i\) is the equivalent calculated with \codename{OClimax} or \codename{Ab2tds}.
Acoustic modes close to the gamma point, which have diverging intensity, and very low intensity data have been excluded to avoid numerical instabilities (more details are in the \supp). The \codename{OClimax} neutron dynamical structure factors have been read directly from \codename{OClimax} output. In the case of \codename{Ab2tds}, the instrumental resolution applied when creating such maps could not be completely removed from the \codename{Ab2tds} output, so the dynamical structure factor has instead been calculated by binning in energy the \codename{Ab2tds} mode-resolved output (which is equivalent to the one-phonon structure factor in \cref{eqn:coh_sfac} with the \(\left( n_{\mathbf{q}\nu} + 1\right)\) factor from \cref{eqn:coh_intensity} already applied). 

The results of the comparisons are summarised in \cref{table:ab2tds} (\codename{Ab2tds}) and \cref{table:oclimax} (\codename{OClimax}). The right-hand column of each table shows the MRPD for \sqw{} calculated from phonon frequencies and eigenvectors obtained from interpolation via \codename{CASTEP} (Nb, quartz, \lzo{}) or \codename{Phonopy} (Al). These MRPD test the computation of \sqw{} by \codename{Euphonic} directly from the frequencies and eigenvectors. The previous column shows the comparison when in the case of \codename{Euphonic} \sqw{} is computed from the force constants, in order to test the interpolation available in \codename{Euphonic} in addition to  the \sqw{} calculation from the phonon frequencies and eigenvectors computed by that interpolation.
For \codename{Ab2tds} the agreement is extremely good, with MRPDs of 0.05\% or less.
In the case of \codename{OClimax}, the MRPDs are significantly larger, up to 2.62\% for the \lzo{} [-5, 7 -L] cut at 300K. However, the overall comparison is still reasonably good, without particularly systematic variation of the relative percentage difference across the slices. The scripts used for validation and the \codename{Euphonic}, \codename{Ab2tds} and \codename{OClimax} inputs and outputs are available at \url{https://github.com/pace-neutrons/euphonic-validation}. Further details on the validation calculations are given in the \supp{}.

%============================================================================================
%============================================================================================
%           Section 6. Examples
%============================================================================================
%============================================================================================

\section{Examples}

\label{section:examples}

\subsection{Command-line tools and plotting}

\label{section:cl_tools}

This section illustrates some of the main features of \codename{Euphonic}, using a variety of samples for both single crystals and powders, from a range of modelling codes. An overview of each of the command-line tools in \codename{Euphonic} is shown in \cref{table:cl_tool_summary}, with example figures referenced for each. Many of these tools allow sampling parameters (e.g. energy bins, broadening) and appearance options (e.g. unit conversions, axis labels and styling) to be easily specified via command-line arguments.

\begin{figure}
\caption{Example \codename{Euphonic} script to calculate and plot the neutron dynamical structure factor in the \([h,-h,-2]\) direction, producing a plot similar to \csubref{fig:validation_cuts}{f}}
\scalebox{1.}{\includegraphics{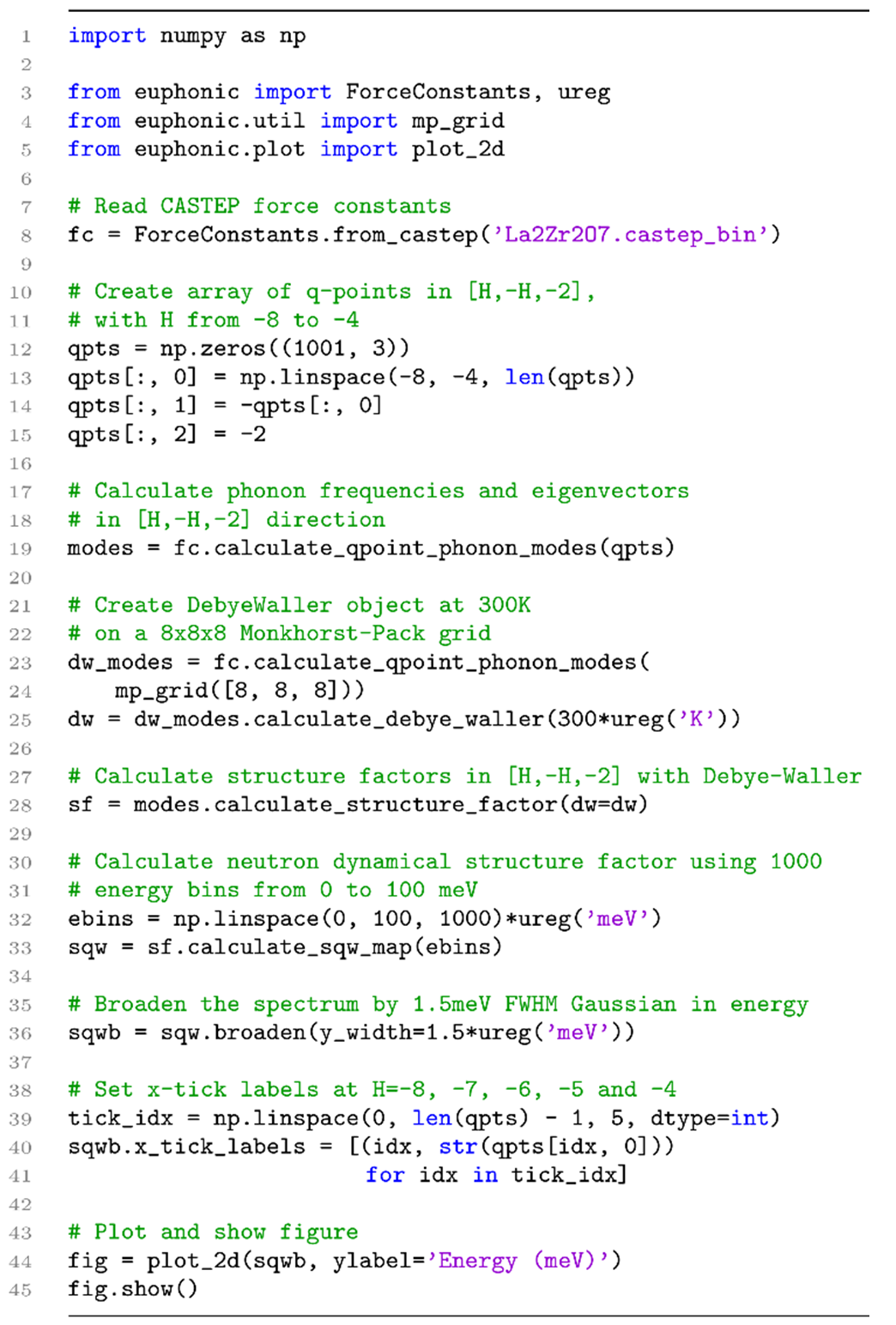}}
\label{fig:code_example}
\end{figure}

For custom plots it may be necessary to write a Python script. For example, \cref{fig:validation_cuts} shows the neutron dynamical structure factor sampled along arbitrary \boldQ{} directions. A sample script to achieve this kind of result is shown in \cref{fig:code_example}.

%============================================================================================

\subsection{Experimental powder data comparison}

\label{section:experimental_data_powder}

\begin{figure}
\caption{Al powder data at 5~K. \sfmt{a} Experimental data recorded on MARI \sfmt{b} Corresponding powder-averaged coherent \smodqw{} generated with \code{euphonic-powder-map}.}
\scalebox{1.}{\includegraphics{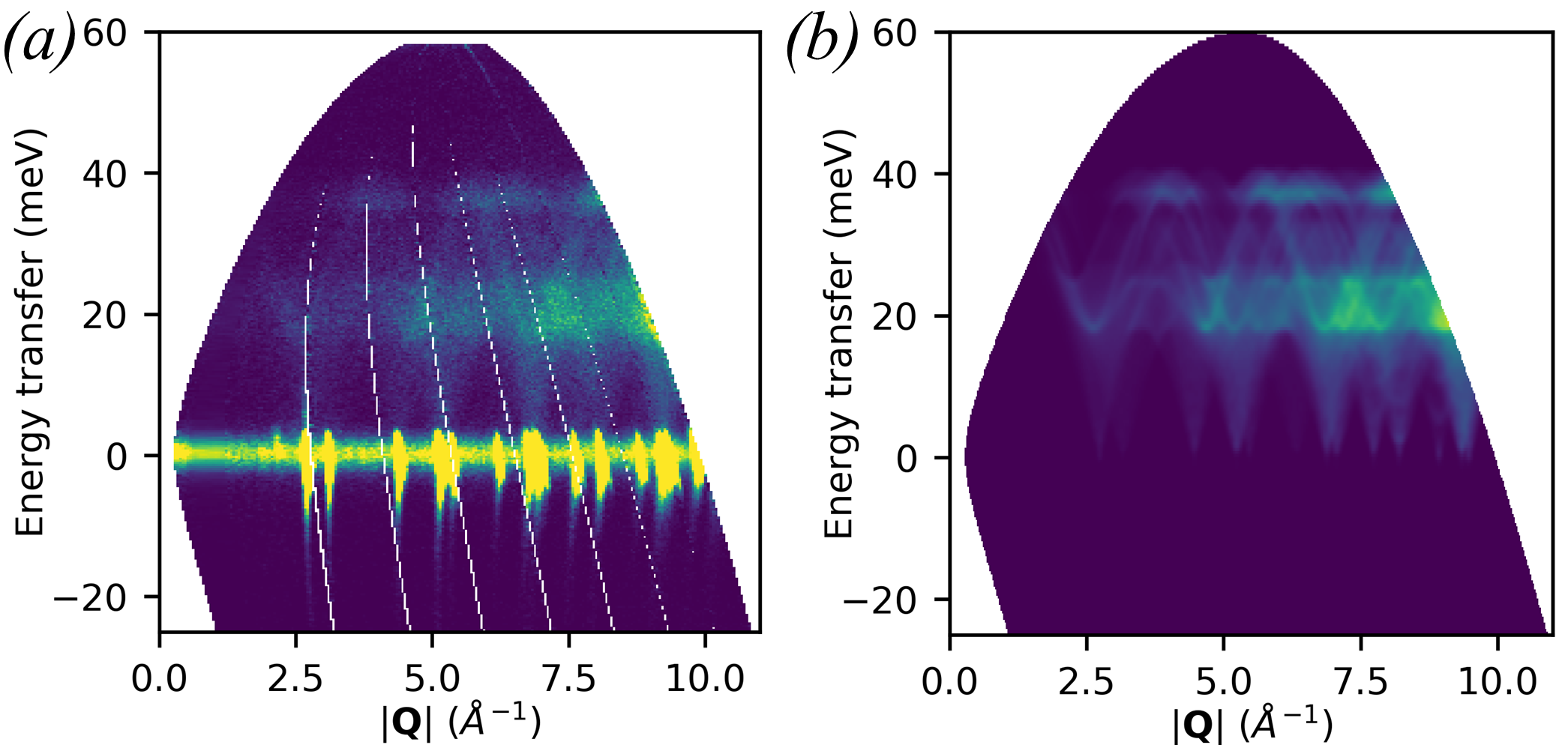}}
\label{fig:al-powder}
\end{figure}

\begin{figure}
\caption{Si powder data at 300~K. \sfmt{a} Experimental data recorded on MARI \sfmt{b} Corresponding powder-averaged coherent \smodqw{} generated with \code{euphonic-powder-map}.}
\scalebox{1.}{\includegraphics{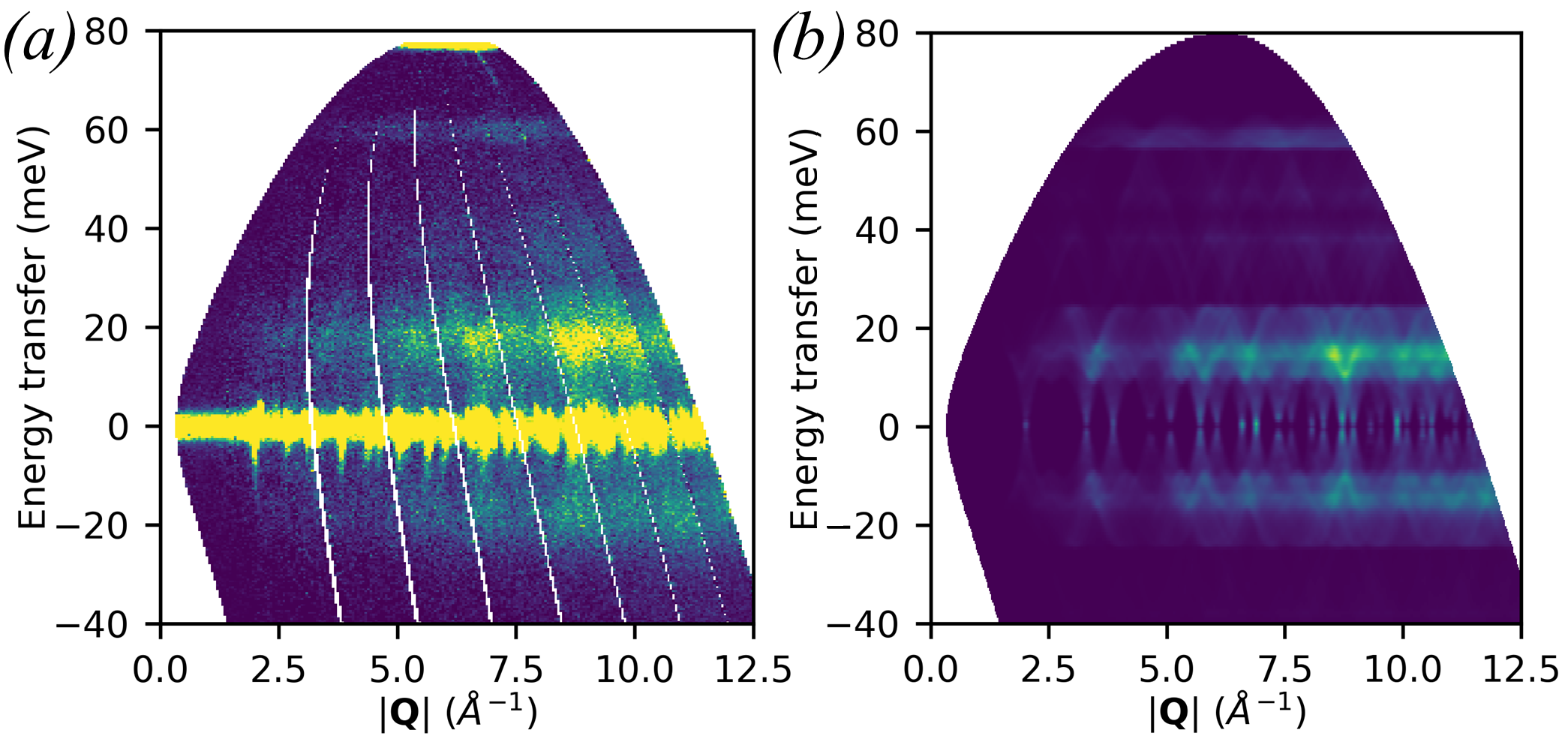}}
\label{fig:si-powder}
\end{figure}

Here, prior experimental measurements are compared with newly simulated spectra~\cite{fair_phonon_2022}.
Measurements of powdered elemental samples of Al (at 5~K with 60~meV incident energy and the Gd monochromating chopper running at 200~Hz) and Si (at 300~K with 80~meV incident energy with the ``sloppy"\footnote{The term ``sloppy" here means the chopper has relative wide openings to maximize flux over a wide range of energies (particularly low energies < 100meV), in contrast to other choppers which are optimized for sharp resolution at specific (high) energies but at the cost of lower transmission and flux.} monochromating chopper at 250~Hz) were recorded on the MARI instrument at ISIS, the data were reduced using \codename{Mantid} and binned to 2D \smodqw{} maps using \codename{MSLICE} in \codename{Mantid} (\csubref{fig:al-powder}{a} and \csubref{fig:si-powder}{a}).
% Comment - MARI Al run number 21335; Si run number 21975
For the Al calculations, \codename{VASP} 5.4.4 was used, and the force constants were obtained by finite displacements using \codename{Phonopy}~\cite{kresse_efficient_1996,kresse_ultrasoft_1999,togo_first_2015}.
For Si, \codename{CASTEP} was used, obtaining force constants by 
 density-functional perturbation theory (DFPT)~\cite{clark_first_2005, refson_variational_2006}. For more details, see the \supp.

The resulting \filename{castep.bin} and \filename{phonopy.yaml} files were used with the \code{euphonic-powder-map} command-line tool to generate numerically sampled maps of \(S_{\mathrm{coh}}(\modQ{}, \omega)\).
(Details of the parameters are given in the \supp).
These results are plotted in \csubref{fig:al-powder}{b} and \csubref{fig:si-powder}{b} and include the main inelastic scattering features in the positive region of the experimental measurements.
It is easy to see in both the experimental and simulated data how spherically averaged periodic features collide to create broad regions of higher intensity. In the case of the 300~K Si data, both the experimental and simulated results have significant intensity in the negative energy transfer region, whereas for the 5~K Al data the intensity is suppressed in this region. This arises from the Bose population factor suppressing the low temperature scattering.
The high intensities seen around zero energy transfer in the experimental measurements in \csubref{fig:al-powder}{a} and \csubref{fig:si-powder}{a} are due to scattering from Bragg peaks which is not currently modelled by Euphonic, hence they are absent from the corresponding simulations in \csubref{fig:al-powder}{b}. and \csubref{fig:si-powder}{b}. The high intensity feature in \csubref{fig:si-powder}{a} at 80 meV is an artefact due to a later pulse of neutrons which passes through the chopper system, and is not due to scattering from the sample, so is not reproduced in the simulated data in \csubref{fig:si-powder}{b}.

%============================================================================================

\subsection{Experimental single crystal data comparison using \codename{Horace}}

\label{section:experimental_data_crystal}

Simulations created from pre-existing quartz force constants calculations have been compared with newly collected experimental data~\cite{fair_phonon_2022}.
Original quartz force constants calculations performed with \codename{CASTEP} 6.1 have been re-processed for this work with \codename{CASTEP} 19.1; more details are in the \supp.
For the experimental measurements, a large single crystal of natural quartz was aligned with the c-axis vertical on an Al plate, secured in place with Al wire. It was cooled to 10 K using a closed cycle refrigerator. The MERLIN ‘G’ chopper at 350 Hz was phased to select 45 meV incident energy neutrons. The sample was rotated over 180 \(^\circ\) in 0.5 \(^\circ\) steps. The data for each individual angle were reduced using \codename{Mantid} and then combined using \codename{Horace}.

\begin{figure}
\caption{\sfmt{a}, \sfmt{c} and \sfmt{e} cuts through the experimental quartz dataset and \sfmt{b}, \sfmt{d} and \sfmt{f} corresponding simulations with \codename{Euphonic}. All figures use the projection axes of \([h,0,0]\), \([\frac{h}{2},-h,0]\), \([0,0,l]\) with integration over the non-plotted directions of \(\pm\) 0.1. The energy axis integration in \sfmt{e} and \sfmt{f} is from 8 to 9 meV.}
\scalebox{1.}{\includegraphics{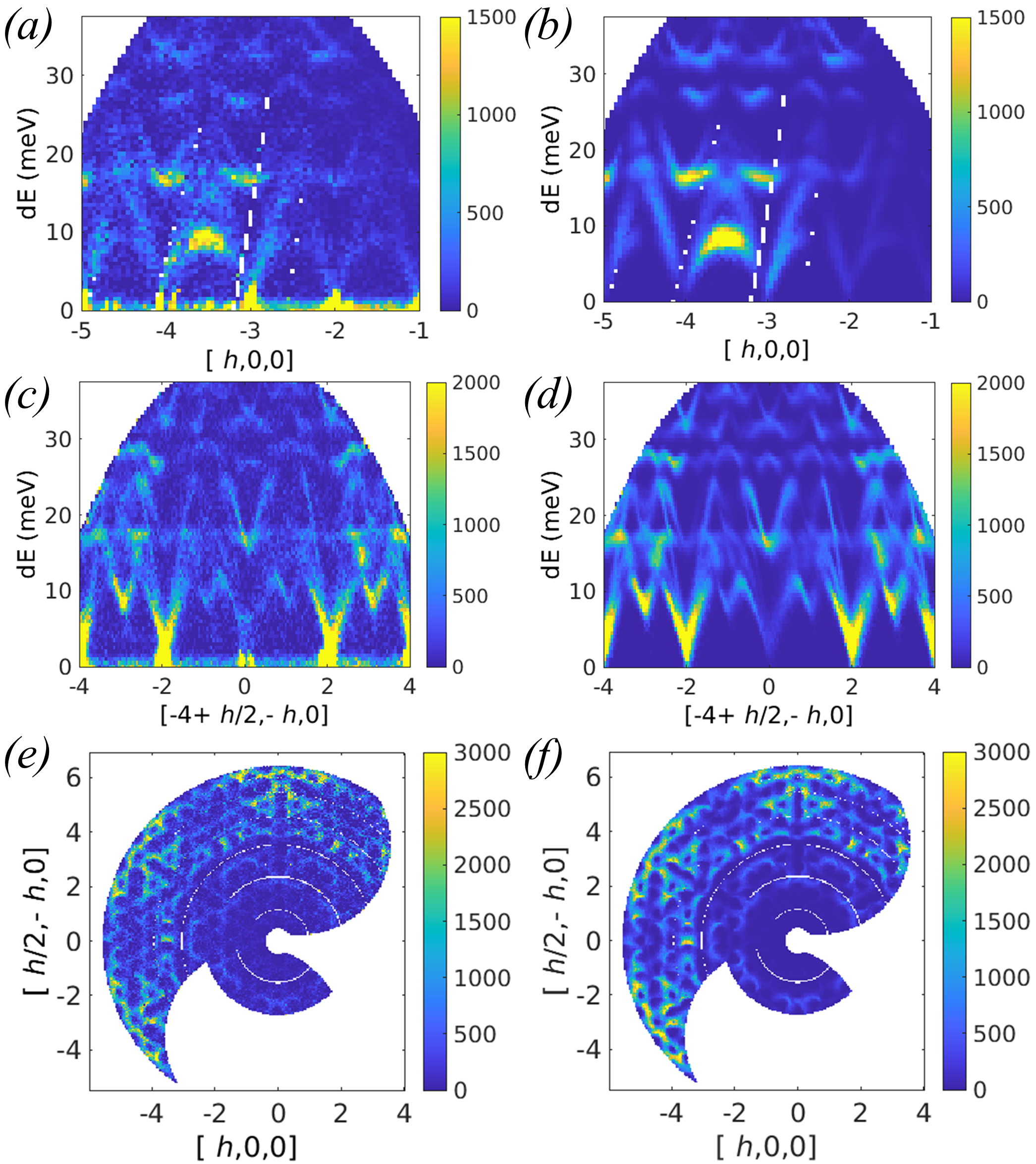}}
\label{fig:quartz_colour_plots}
\end{figure}

\begin{figure}
\caption{Linecuts through the quartz data and corresponding simulations with \codename{Euphonic} (solid lines). \sfmt{a} a cut through [-3,0,0], \sfmt{b} [-5,1,0], \sfmt{c} [-3.75,-0.5,0], and \sfmt{d} a cut at constant energy (integrated from 8 to 9 meV). All figures use the projection axes of \([h,0,0]\), \([\frac{h}{2},-h,0]\), \([0,0,l]\) with integration over the non-plotted directions of \(\pm\) 0.1.}
\scalebox{1.}{\includegraphics{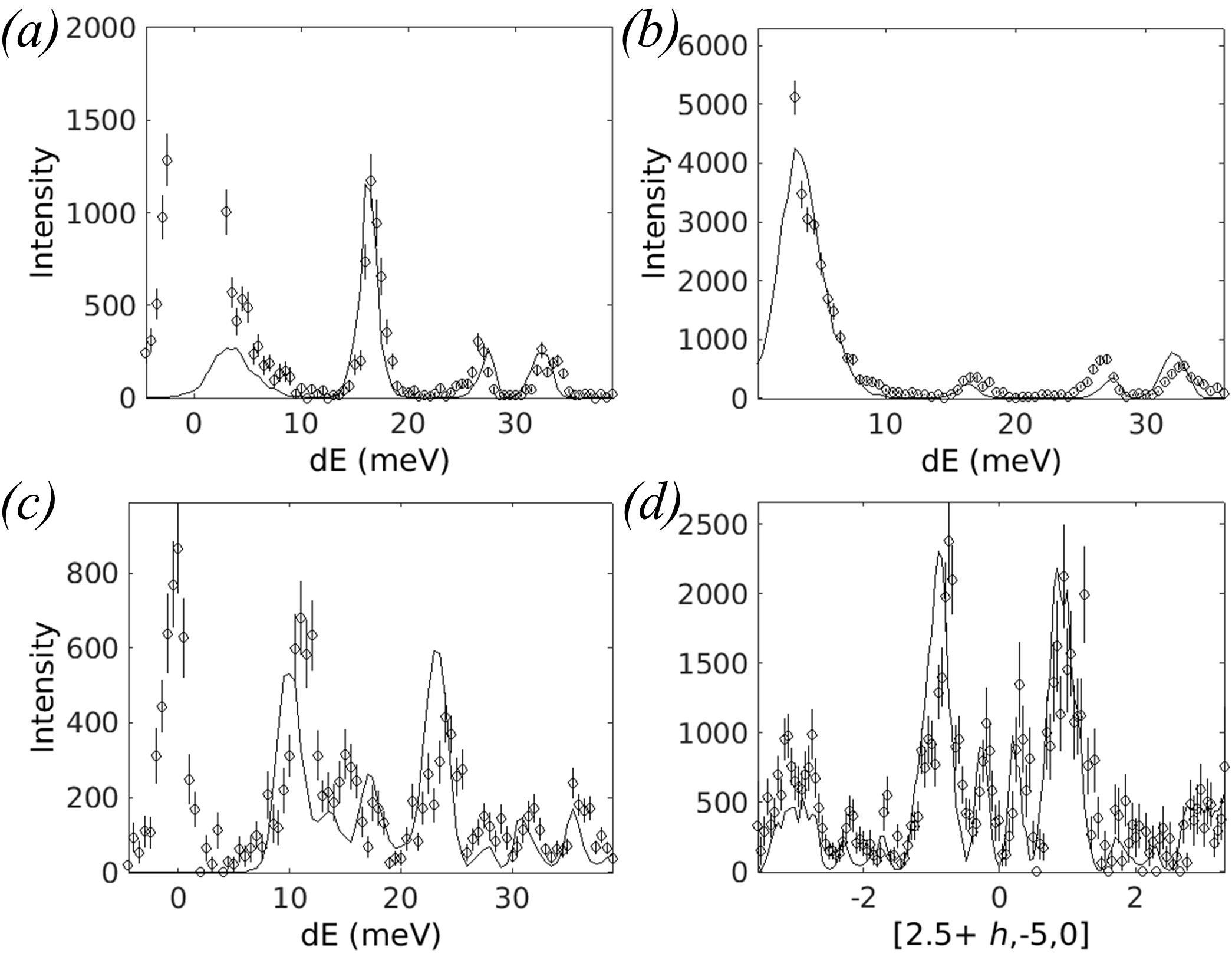}}
\label{fig:quartz_line_plots}
\end{figure}

Several cuts through experimental and simulated data for quartz are shown in \cref{fig:quartz_colour_plots}. The neutron dynamical structure factor has been computed with \codename{Euphonic}, with the instrumental resolution accounted for by the Monte Carlo resolution convolution method \cite{perring_high_1991} implemented in \codename{Horace}. In addition, 1D cuts are shown in \cref{fig:quartz_line_plots}, with the simulated scattering likewise convolved with the instrumental resolution function. The only adjustable parameter in the comparison is a global scaling factor which has been determined by setting the integrated areas of the experimental and simulated data in \csubref{fig:quartz_line_plots}{a} between 12 and 37 meV to be equal, which has then been applied to all 1D and 2D cuts. Aside from this (arbitrary) choice of intensity scale, no adjustable parameters have been used in any part of the calculation. There are small disagreements, for example the phonon frequencies do not match up perfectly. This is to be expected, as the LDA tends to overestimate bond strengths (and thus phonon frequencies). In addition there are small differences in particular mode intensities, for example around 27 meV in \csubref{fig:quartz_line_plots}{b} where the intensity is underestimated or at 23 meV in \csubref{fig:quartz_line_plots}{c} where it is overestimated. Given the lack of adjustable parameters however the agreement is remarkably strong. Notwithstanding disagreements that arise from the \abinitio{} code not fully capturing the physics of the material, \csubref{fig:quartz_line_plots}{a-d} show the importance of accounting for instrumental resolution when comparing calculation with experimental data.

%============================================================================================
%============================================================================================
%           Section 7. Conclusions
%============================================================================================
%============================================================================================

\section{Conclusions and future development}

We have described the \codename{Euphonic} package which is designed to efficiently compute phonon eigenvectors, eigenvalues and the inelastic neutron scattering cross-section for a large number of \qpts{} from force constants matrices. It has a set of command lines tools to plot the phonon band structure, DOS and the neutron dynamical structure factor along a path in reciprocal space, and an extensive Python API so that it can be used directly from Python environments for customised workflows and plotting. \codename{Euphonic} also works directly from \codename{Horace} \cite{ewings_horace_2016} and the \codename{Mantid}~\cite{arnold_mantiddata_2014} plug-in \codename{AbINS}~\cite{dymkowski_abins_2018}.

Examples of the use of the \codename{Euphonic} command line tools were shown in \cref{section:examples}, together with comparison of simulated and experimental data from powders and single crystals, the latter also including convolution with the instrumental resolution function.

\codename{Euphonic} has been extensively benchmarked and validated against other codes. It is now being used for the interpretation of phonon data at the ISIS Neutron and Muon Source, where it will continue to be maintained and developed as part of the core data analysis software portfolio. Promising avenues for future development include: support of other codes, particularly the \codename{Atomic Simulation Environment} (ASE)~\cite{hjorth_larsen_atomic_2017}; performance improvements from symmetry-aware interpolation of the dynamical matrices, and other interpolation methods; corrections for thermal expansion/softening by using a quasi-harmonic force-constants matrix; and X-ray structure factors.

\codename{Euphonic} is open-source and the source code is available to download from Github~\cite{fair_euphonic_2022}, with releases also available via the \codename{pip} and \codename{conda} package managers, as a service to the neutron scattering and the computational chemistry and physics communities. The authors welcome bug reports, feedback on usability and documentation, and suggestions for additional functionality. The authors also welcome contributions to the \codename{Euphonic} package.

     % Appendices appear after the main body of the text. They are prefixed by
     % a single \appendix declaration, and are then structured just like the
     % body text.

%\appendix

     %-------------------------------------------------------------------------
     % The back matter of the paper - acknowledgements and references
     %-------------------------------------------------------------------------

     % Acknowledgements come after the appendices

\section*{Acknowledgments}
The computing resources for \cref{section:performance} were provided by STFC Scientific Computing Department’s SCARF cluster.

     % References are at the end of the document, between \begin{references}
     % and \end{references} tags. Each reference is in a \reference entry.
     
%\begin{references}
%\reference{Author, A. \& Author, B. (1984). \emph{Journal} \textbf{Vol}, 
%first page--last page.}
%\end{references}

\printbibliography

     %-------------------------------------------------------------------------
     % TABLES AND FIGURES SHOULD BE INSERTED AFTER THE MAIN BODY OF THE TEXT
     %-------------------------------------------------------------------------

     % Simple tables should use the tabular environment according to this
     % model

%\begin{table}
%\caption{Caption to table}
%\begin{tabular}{llcr}      % Alignment for each cell: l=left, c=center, r=right
% HEADING    & FOR        & EACH       & COLUMN     \\
%\hline
% entry      & entry      & entry      & entry      \\
% entry      & entry      & entry      & entry      \\
% entry      & entry      & entry      & entry      \\
%\end{tabular}
%\end{table}

     % Postscript figures can be included with multiple figure blocks

%\begin{figure}
%\caption{Caption describing figure.}
%\includegraphics{fig1.ps}
%\end{figure}

\onecolumn

%=============================================================================================
\begin{table}
\caption{\codename{CASTEP} \filename{.phonon} file size, and time taken to write the \filename{.phonon} file compared with the total time taken to run the \codename{CASTEP} \code{phonons} tool (including write time) with 25,000 \qpts{} for different materials. (Details of the hardware are given in \cref{section:hardware_and_software}.)}
\vspace{5mm}
\begin{tabular}{llrrr}
 Material & Atoms & Size (GB) & Total Time (s) & Write Time (s) \\
 \hline
 Nb & 1 & 0.025 & 57.837 & 7.557 \\
 Quartz & 9 & 0.913 & 811.430 & 64.930 \\
 \lzo{} & 22 & 5.187 & 302.167 & 258.243 \\
\end{tabular}
\label{table:castep_phonons_write}
\end{table}

%=============================================================================================
\begin{table}
\caption{Comparison of the mean time taken for phonon interpolation against the mean time taken to calculate the mode-resolved structure factors for 25,000 \qpts{} with serial Python in \codename{Euphonic}. (Details of the hardware are given in \cref{section:hardware_and_software}.)}
\vspace{5mm}
\begin{tabular}{lrr}
 Material & Interpolation (s) & Structure Factor (s) \\
 \hline
 Nb & 7.070 & 0.104 \\
 Quartz & 199.069 & 0.426 \\
 \lzo{} & 35.743 & 1.853 \\
\end{tabular}
\label{table:compare_interpolate_sf}
\end{table}

%=============================================================================================

%=============================================================================================

%=============================================================================================

%=============================================================================================

%=============================================================================================

%=============================================================================================

%=============================================================================================
\begin{table}
\caption{Mean relative percentage difference between the \codename{Euphonic} and \codename{Ab2tds} neutron dynamical structure factor for the Nb, quartz and \lzo{} 2D (\boldQ,\(\omega\)) maps shown in \cref{fig:validation_cuts} at 300K. For \codename{Euphonic} the neutron dynamical structure factor has been calculated from both \codename{CASTEP}-interpolated and \codename{Euphonic}-interpolated frequencies and eigenvectors.}
\vspace{5mm}
\begin{tabular}{lcrr}
 \multirow{3}{*}{Material} &
 \multirow{3}{*}{\boldQ -direction} &
 \multicolumn{2}{c}{Mean Relative Percentage Difference} \\
 && \codename{Euphonic} Interpolation &
    \codename{CASTEP} Interpolation \\
 \hline
 \multirow{2}{*}{Nb} &
  {\([h,h,0]\)} & \textless 0.01 & \textless 0.01 \\
  & {\([2-k,k,0]\)} & \textless 0.01 & \textless 0.01 \\
 \hline
 \multirow{2}{*}{Quartz} &
  {\([h,-4,0]\)} & 0.03 & 0.03 \\
  & {\([-3,0,-l]\)} & 0.05 & \textless 0.01 \\
 \hline
 \multirow{2}{*}{\lzo} &
  {\([-5,7,-l]\)} & 0.05 & 0.05 \\
  & {\([h,-h,2]\)} & 0.05 & 0.05 \\
\end{tabular}
\label{table:ab2tds}
\end{table}

%=============================================================================================
\begin{table}
\caption{Mean relative percentage difference between the \codename{Euphonic} and \codename{OClimax} neutron dynamical structure factor for the 2D (\boldQ,\(\omega\)) maps shown in \cref{fig:validation_cuts} at different temperatures. For \codename{Euphonic} the neutron dynamical structure factor has been calculated from both \codename{CASTEP}/\codename{Phonopy}-interpolated ((Nb, quartz and \lzo{})/Al) and \codename{Euphonic}-interpolated frequencies and eigenvectors.}
\vspace{5mm}
\begin{tabular}{lccrr}
 \multirow{3}{*}{Material} &
 \multirow{3}{*}{\boldQ -direction} &
 \multirow{3}{*}{T (K)} &
 \multicolumn{2}{c}{\scriptsize Mean Relative Percentage Difference} \\
 &&& \codename{Euphonic} &
     \codename{CASTEP}/\codename{Phonopy} \\
 &&& Interpolation & Interpolation \\
 \hline
 \multirow{4}{*}{Nb} &
   \multirow{2}{*}{\([h,h,0]\)} &
    300 & \textless 0.01 & \textless 0.01 \\
    && 5 & \textless 0.01 & \textless 0.01 \\
 & \multirow{2}{*}{\([2-k,k,0]\)} &
    300 & \textless 0.01 & \textless 0.01 \\
    && 5 & \textless 0.01 & \textless 0.01 \\
 \hline
 \multirow{4}{*}{Quartz} &
   \multirow{2}{*}{\([h,-4,0]\)} &
    300 & 0.87 & 0.87 \\
    && 5 & 0.49 & 0.49 \\
 & \multirow{2}{*}{\([-3,0,-l]\)} &
    300 & 1.75 & 1.82 \\
    && 5 & 0.83 & 0.92 \\
 \hline
 \multirow{4}{*}{\lzo} &
    \multirow{2}{*}{\([-5,7,-l]\)} &
    300 & 2.62 & 2.62 \\
    && 5 & 1.83 & 1.83 \\
  & \multirow{2}{*}{\([h,-h,2]\)} &
    300 & 2.05 & 2.05 \\
    && 5 & 1.42 & 1.42 \\
 \hline
 \multirow{4}{*}{Al} &
   \multirow{2}{*}{\([h,2,2]\)} &
    300 & 0.01 & \textless 0.01 \\
    && 5 & 0.01 & \textless 0.01 \\
  & \multirow{2}{*}{\([h,2+\frac{h}{2},2+\frac{h}{2}]\)} &
    300 & \textless 0.01 & \textless 0.01 \\
    && 5 & \textless 0.01 & \textless 0.01 \\
\end{tabular}
\label{table:oclimax}
\end{table}
%=============================================================================================
\begin{table}
\caption{An overview of the  command-line tools available in \codename{Euphonic}}
%\begin{tabular}{lll}
\vspace{5mm}
\begin{tabular}{p{5.5cm}p{8cm}p{3cm}}
 Tool & Description & Example Figure \\
 \hline
 \code{euphonic-dispersion} &
 Plots a phonon band structure along a recommended \qpt{} path (generated by \codename{SeeK-path}\cite{hinuma_band_2017})  if using a force constants file as input, or will plot existing phonon frequencies if using a phonon modes file as input (e.g. \codename{CASTEP} \filename{.phonon}) file & \csubref{fig:cl_tool_plots}{a} \\
 \hline
 \code{euphonic-dos} &
 Plots a total or partial DOS on a specified Monkhorst--Pack grid if using a force constants file, or on a precalculated grid if using a phonon modes file. The spectrum can also be weighted by coherent or incoherent neutron-scattering cross-section &
 \csubref{fig:cl_tool_plots}{d} \\
 \hline
 \code{euphonic-intensity-map} &
 Plots a 2D \((\boldQ,\omega)\) crystal intensity map along a recommended \qpt{} path if using force constants, or along a precalculated path if using phonon modes. The intensities can be weighted by the neutron dynamical structure factor or phonon DOS &
 \csubref{fig:cl_tool_plots}{b} \& \csubref{fig:cl_tool_plots}{c} \\
 \hline
 \code{euphonic-powder-map} &
 Plots a 2D \((\modQ,\omega)\) powder intensity map along a specified range in \modQ{} using spherical averaging; requires force constants. Intensities can be weighted by the neutron dynamical structure factor or phonon DOS &
 \csubref{fig:al-powder}{b} and \csubref{fig:si-powder}{b} \\
 \hline
 \code{euphonic-show-sampling} &
 Plots a 3D visualisation of the distribution of \qpts{} over a sphere for the different spherical sampling schemes that are used in powder averaging &
 \cref{fig:show-sampling-golden-sphere} \\
 \hline
  \code{euphonic-optimise-dipole-parameter} &
 The 'dipole parameter' determines the balance of real and reciprocal terms used in the Ewald sum for calculating the dipole correction (see \cref{subsection:polar_crystals}). A higher value uses more reciprocal terms, and a lower value more real terms. Tuning this parameter can improve performance; this tool runs the interpolation for a few \qpts{} for a few different values of the parameter, and suggests an optimum value &
 - \\
\end{tabular}
\label{table:cl_tool_summary}
\end{table}

%=============================================================================================
\begin{figure}
\caption{Examples of plots produced with command-line tools in \codename{Euphonic}. \sfmt{a} Quartz with \code{euphonic-dispersion}. \sfmt{b} Quartz with \code{euphonic-intensity-map} using the DOS-weighted intensities option. \sfmt{c} Quartz with \code{euphonic-intensity-map} using the neutron dynamical structure factor weighted intensities option. \sfmt{d} \lzo{} with \code{euphonic-dos} using the coherent-weighted partial DOS option.}
\scalebox{1.}{\includegraphics{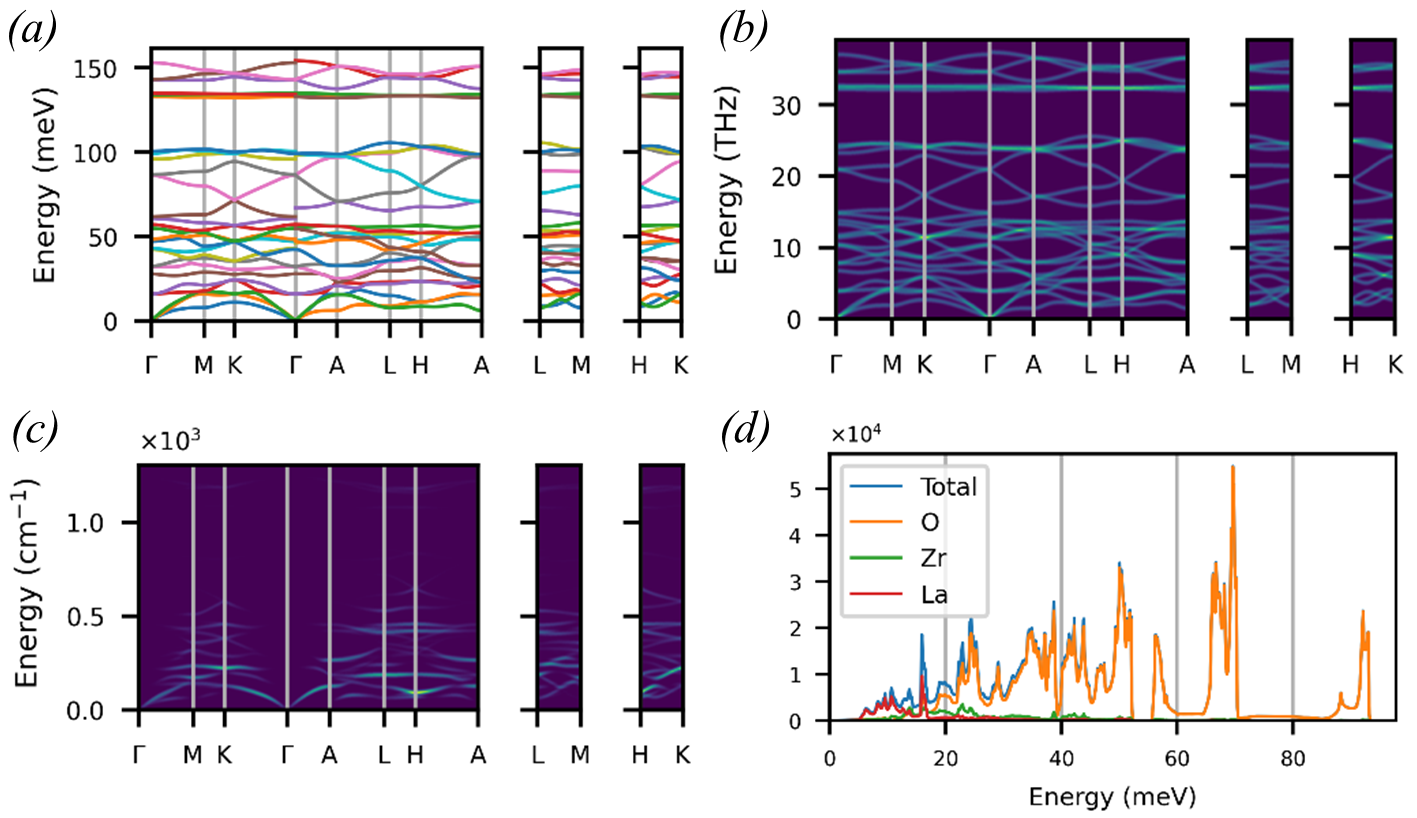}}
\label{fig:cl_tool_plots}
\end{figure}

%=============================================================================================
\begin{figure}
\caption{Outputs of \code{euphonic-show-sampling}. \sfmt{a} "Golden sphere" sampling \sfmt{b} "Improved spherical polar" sampling with jitter.}
\scalebox{1.}{\includegraphics{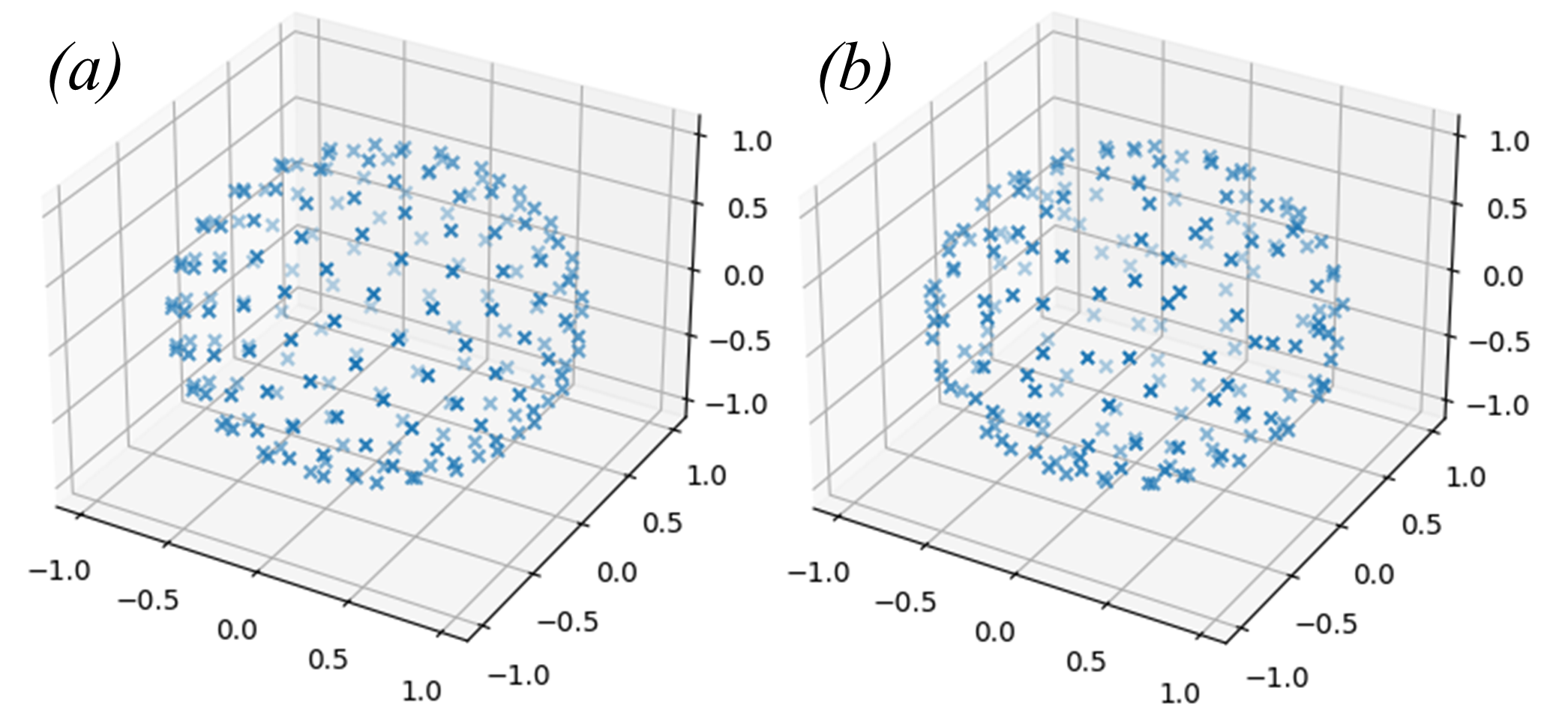}}
\label{fig:show-sampling-golden-sphere}
\end{figure}

%=============================================================================================

%=============================================================================================

%=============================================================================================

%=============================================================================================

%=============================================================================================

\twocolumn

\end{document}

\title{Supporting information for:\\ \codename{Euphonic}: inelastic neutron scattering simulations from force constants and visualisation tools for phonon properties}
\author[a]{Rebecca Fair}
\author[b]{Adam Jackson}
\author[a,c]{David Voneshen}
\author[b]{Dominik Jochym}
\author[a]{Duc Le}
\author[a]{Keith Refson}
\author[a]{Toby Perring}

\affil[a]{ISIS Neutron and Muon Source, STFC Rutherford Appleton Laboratory, Harwell Campus, Didcot, Oxfordshire OX11 0QX, UK}
\affil[b]{Scientific Computing Department, STFC Rutherford Appleton Laboratory, Harwell Campus, Didcot, Oxfordshire OX11 0QX, UK}
\affil[c]{Department of Physics, Royal Holloway University of London, Egham, TW20 0EX, UK}
\date{}
\maketitle
\onecolumn

\section{Validation calculation}
\label{section:validation_supplementary}

\subsection{Intensity scaling}

For the comparison of \codename{Euphonic}~\cite{fair_euphonic_2022} with \codename{Ab2tds}~\cite{mirone_ab2tds_2013} and \codename{OClimax}~\cite{cheng_simulation_2019}, a scaling factor was calculated for each intensity map by averaging the ratio of the intensities in each bin with those in the corresponding bins in the \codename{Euphonic} intensity map. Data points at energies less than 1~meV were first removed to eliminate acoustic mode data close to the gamma points (where the phonon intensity diverges) and afterwards any values that were smaller than $10^{-4}$ of the maximum intensity in the remaining portion of the map were removed to avoid potential numerical instabilities arising in the computation of very small intensities. The \codename{Ab2tds} or \codename{OClimax} intensity map was then normalised by the resulting scaling factor and the mean relative percentage difference (MRPD) was computed for the data that passed the same filtering criteria.

\subsection{Symmetrisation of the Debye--Waller exponent}

In all cases the grid used to calculate the Debye--Waller factor was a fully unfolded Monkhorst--Pack grid, i.e. the set of points was not symmetry reduced. It was found that using a symmetry reduced grid produces slightly different values for the Debye--Waller exponent \(W_\kappa\) than the full grid unless the Debye--Waller exponent is explicitly symmetrised so that it is invariant under symmetry operations. This would have made comparison between the different codes difficult as \codename{Ab2tds} and \codename{OClimax} appear to handle symmetry reduced grids differently, increasing the MRPDs depending on which grid type and symmetrisation settings are used. (\codename{Euphonic} has the option to switch on or off Debye--Waller symmetrisation).
An example of the MRPD increase for \codename{Ab2tds} is shown in Table 1 for the case where a symmetry reduced grid was used as input and Debye--Waller symmetrisation was disabled in \codename{Euphonic}. The MRPDs are higher than in the corresponding \cref{table:ab2tds} in the main text, where a full grid was used. If the Debye--Waller symmetrisation is enabled again in \codename{Euphonic} when using the reduced grid, the low MRPDs seen in \cref{table:ab2tds} in the main text are recovered. 

\begin{table}[h]
\label{table:ab2tds_dw}
\caption{Mean relative percentage difference between \codename{Euphonic} and \codename{Ab2tds} as in \cref{table:ab2tds}
in the main text, but using a symmetry reduced Monkhorst--Pack grid to calculate the Debye--Waller factor, and with Debye--Waller symmetrisation turned off in \codename{Euphonic}}
\begin{tabular}{lccc}
\multirow{3}{*}{Material} &
\multirow{3}{*}{\boldQ -direction} &
\multicolumn{2}{c}{Mean Relative Percentage Difference} \\
 && \codename{Euphonic} Interpolation &
    \codename{CASTEP} Interpolation \\
\hline
\multirow{2}{*}{Nb} &
 {\([h,h,0]\)} & 0.21 & 0.21 \\
 & {\([2-k,k,0]\)} & 0.25 & 0.25 \\
\hline
\multirow{2}{*}{Quartz} &
 {\([h,-4,0]\)} & 1.89 & 1.89 \\
 & {\([-3,0,-l]\)} & 0.27 & 0.23 \\
\hline
\multirow{2}{*}{\lzo} &
 {\([-5,7,-l]\)} & 1.61 & 1.61 \\
 & {\([h,-h,-2]\)} & 1.15 & 1.15 \\
\end{tabular}
\end{table}

\section{Force constants calculation details}

The input and output files for each of the following calculations are available at \url{https://doi.org/10.5281/zenodo.6620084}~\cite{fair_phonon_2022}.

\subsection{\lzo{}}
Calculations of the force constants were performed with DFT as implemented in \codename{CASTEP} 17.21 \cite{clark_first_2005}. Default ultrasoft pseudopotentials were used and the results presented used the Local Density Approximation (LDA) to exchange and correlation. A plane wave cut-off of 600 eV with electronic \kpt{} sampling on a Monkhorst--Pack mesh of $4 \times 4 \times 4$ within the primitive cell was found to reduce the error in the forces to below 0.005 eV~Å$^{-1}$. The structure and lattice parameters were relaxed with the quasi-Newton method \cite{byrd_representations_1994} corrected for the finite basis set \cite{francis_finite_1990}. Force constants were calculated using the finite displacement/supercell method \cite{frank_ab_1995} within a single cubic unit cell. 

\subsection{Quartz}
Original calculations were performed with \codename{CASTEP} 6.1 using LDA exchange and correlation, optimised norm-conserving pseudopotentials~\cite{rappe_optimized_1990}, a plane-wave cutoff of 880 eV and a $5 \times 5 \times 4$ Monkhorst--Pack mesh for \kpt{} sampling. The optimised lattice parameters were 4.852\AA\ and 5.350\AA. DFPT calculations also sampled using a $5 \times 5 \times 4$ grid with dipole-dipole model corrections applied. The resulting force constants were re-processed for this work using \codename{CASTEP} 19.1.

\subsection{Nb}
Niobium calculations used the one-atom primitive cell with bcc lattice parameter of 3.25988\AA, \codename{CASTEP} 19.1, LDA exchange and correlation and a plane-wave cutoff of 900 eV. The NCP19 on-the-fly (OTF) library pseudopotential used is constructed with unfrozen 4s and 4p semi-core states. Electronic states were sampled using an $18 \times 18 \times 18$ Monkhorst--Pack grid and smeared using Gaussian broadening with a width of $0.5$ eV. DFPT calculations were performed for a $9 \times 9 \times 9$ grid of \boldq-points and the force constant matrix constructed by Fourier transforming on this grid.

\subsection{Al}
Force constants were obtained from a $4\times4\times4$ supercell of the cubic (conventional) unit cell using \codename{Phonopy}~\cite{togo_first_2015} to implement and process a symmetrised finite-displacement method. 
The force calculator was \codename{VASP} 5.4.4~\cite{kresse_efficient_1996}, with the standard LDA PAW setup and parameters:
\begin{verbatim}
 ENCUT = 600.000000
 KSPACING = 0.100000
 SIGMA = 0.200000
 EDIFF = 1.00e-08
 ALGO = fast
 PREC = accurate
 IBRION = -1
 ISMEAR = 0
 ISYM = 2
 KPAR = 6
 NCORE = 6
 ADDGRID = .TRUE.
 LASPH = .TRUE.
 LREAL = .FALSE.
\end{verbatim}
After \kpt{} convergence and optimisation the unit cell has a lattice parameter of $\SI{3.984207}{\angstrom}$.

\subsection{Si}

Force constants were obtained from a $5\times5\times5$ $\Gamma$-centered \qpt{} mesh sampling of a Si primitive cell,
using DFPT with \codename{CASTEP} 19.1, on-the-fly-generated norm-conserving pseudopotentials and LDA exchange-correlation functional with a "Precise" (266.9 eV) plane-wave basis set.
The geometry was obtained by local optimisation in \codename{CASTEP} from an initial structure generated with \codename{Atomic Simulation Environment} (ASE)~\cite{hjorth_larsen_atomic_2017}.

\section{Powder-averaged calculation details and parameters}

\subsection{Al}

The powder-averaging command was 
\begin{verbatim}
euphonic-powder-map Al-444-lda.yaml --weighting coherent \
    --q-max 11 --asr --temperature 5 \
    --npts-density 400 --npts-min 500 --npts-max 20000 \
    --energy-broadening 1 --q-broadening 0.1 \
    --e-min -25 --e-max 60 --v-max 2 \
    --angle-range 3 135 --e-incident 60 \
    --no-widgets --style custom.mplstyle \
    --save-to al-simulated.png \
    --x-label '$|\mathbf{Q}|$ $(\AA^{-1})$' \
    --y-label 'Energy transfer (meV)'
\end{verbatim}

\subsection{Si}

The powder-averaging command was 
\begin{verbatim}
euphonic-powder-map Si-prim-555.json --weighting coherent \
    --q-max 12.5 --asr --temperature 300 \
    --npts-density 400 --npts-min 500 --npts-max 20000 \
    --energy-broadening 1 --q-broadening 0.1 \
    --e-min -40 --e-max 80 --v-max 5 \
    --angle-range 3 135 --e-incident 80 \
    --no-widgets --style custom.mplstyle \
    --save-to si-simulated.png \
    --x-label '$|\mathbf{Q}|$ $(\AA^{-1})$' \
    --y-label 'Energy transfer (meV)'
\end{verbatim}

\printbibliography